\documentclass[12pt]{article}
\usepackage{array}
\usepackage{needspace}
\usepackage{float}
\usepackage{tabularx}
\usepackage{amsmath,amssymb,amsfonts,mathrsfs,bbm,epsfig,cancel,cite,setspace,bigstrut,bm,framed,eufrak,tensor,mathtools}
\usepackage[all]{xy}
\usepackage{color}
\usepackage{pifont}
\usepackage{xcolor}
\usepackage{tikz}
\usepackage{mathrsfs}
\usetikzlibrary{positioning}

\usepackage[T1]{fontenc}
\usepackage{lmodern}
\usepackage{amsmath,amssymb,mathtools,bm}
\usepackage{booktabs,tabularx,array}

\usepackage{mathtools}

\newcommand{\dsl}{\pa \kern-0.5em /}

\newcommand{\pa}{\partial}

\newcommand{\bit}{\begin{itemize}}
\newcommand{\eit}{\end{itemize}}

\newcommand{\order}{\mathcal O}

\newcommand{\bea}{\begin{eqnarray}}
\newcommand{\eea}{\end{eqnarray}}
\newcommand{\be}{\begin{equation}}
\newcommand{\ee}{\end{equation}}
\newcommand{\bbe}{\mathbbm{e}}  

\newcommand{\ba}{\begin{array}}
\newcommand{\ea}{\end{array}}

\makeatletter \@addtoreset{equation}{section} \makeatother

\author{JQ}

\newcommand{\cN}{{\cal N}}

\newcommand{\cG}{{\cal G}}

\newcommand{\cH}{{\cal H}}

\newcommand{\cL}{{\cal L}}

\newcommand{\cF}{{\cal F}}
\newcommand{\cA}{{\cal A}}

\newcommand{\cV}{{\cal V}}

\newcommand{\bi}{\begin{itemize}}
\newcommand{\ei}{\end{itemize}}
\newcommand{\beq}{\begin{equation}}
\newcommand{\eeq}{\end{equation}}

\newcommand{\cP}{\mathcal P}

\renewcommand{\thefootnote}{\fnsymbol{footnote}}
\begin{document}

\begin{titlepage}
\vfill
\begin{flushright}
{\tt\normalsize KIAS-P26084}\\

\end{flushright}
\vfill
\begin{center}
{\Large\bf  Self-Force and Bremsstrahlung \\ \vskip 3mm in Finite Electrodynamics }

\vskip 1cm

Kyungsun Lee
and Piljin Yi

\vskip 5mm
{\it School of Physics, Korea Institute
for Advanced Study, Seoul 02455, Korea}

\end{center}
\vfill

\begin{abstract}
\noindent
We study the self-force and the accompanying Bremsstrahlung for a moving
point charge in nonlinear electrodynamics with finite self-field and  
self-energy. The self-force in the Maxwell theory has been the subject of 
many studies, often by renowned physicists. After reviewing 
Coleman's lucid treatise for Maxwell among them, we wish to explore the same question
in the Born--Infeld theory, our prototype. We independently compute the local
self-force and the global radiative dissipation, and show how the two are
mutually consistent. Despite the nonlinearity turning off for
the asymptotic radiative field, its residual effect survives via the
effective multipole moments of a moving charge, allowing deviation from
Maxwell. We close with a detailed discussion about two different manifestations
of the self-force in the literature and how these enter our computation. 
One of the appendices explores Born--Infeld ModMax as a generalization.

\end{abstract}

\vfill
\end{titlepage}

\renewcommand{\thefootnote}{\arabic{footnote}}
\setcounter{footnote}{0}

\parskip 0.2 cm

\section{Born--Infeld versus Maxwell}
\label{sec:bi-maxwell}

The Born--Infeld theory is an invention of the early 20th century, when physicists began
to accommodate point charges such as electrons. One obvious problem with
such a point charge in Maxwell theory is that its electrostatic field comes with
divergent energy density $\sim 1/r^4$ at the origin, so taken literally this means
infinite inertia regardless of the actual mass of the electron. This oddity was later
recognized as the tip of the iceberg that would be eventually resolved by renormalization
of quantum field theory. In the absence of the latter insight, it is only natural
to consider a finite alternative, leading to the Born--Infeld (BI) electrodynamics~\cite{BornInfeld}.
Although BI theory would be later revived in the context of string theory in a new
reincarnation known as Dirac--Born--Infeld (DBI),\footnote{References for DBI would be
too numerous to list; we refer readers to the textbook by Polchinski~\cite{Polchinski} who had discovered the
D-branes whose worldvolume dynamics proved to be supersymmetric DBI.} it was by and large forgotten by
the mainstream community once the quantum field theory framework arrived and settled
in as the main toolkit.

The divergent behavior of the Maxwellian self-energy led to other interesting
consequences, a classic example of which is the self-force that accounts for the
dissipative nature of Bremsstrahlung. The latter carries away energy flux proportional
to acceleration squared, and the energy budget requires this energy to be extracted
from the particle's kinetic energy. This has led to the Abraham--Lorentz--Dirac (ALD)
damping force~\cite{Abraham1903,Abraham1905,Lorentz1909,Dirac1938} on the moving 
charge as the only possible explanation, proportional to the
time derivative of the acceleration, yet it took decades of research to understand  
how this self-force comes about from a first-principle computation. One prime puzzle was
the case of a uniformly accelerated charge, whose damping force vanishes yet
a constant outflux of energy persists. The latter puzzle was resolved by Boulware's
definitive work~\cite{Boulware}, absolving the ALD self-force handsomely.

Proper ``ab initio'' derivation of the ALD damping force as a self-force of the moving charge on
itself has an involved history and was addressed by many renowned physicists~\cite{WheelerFeynman,DeWittBrehme,Coleman,Rohrlich1961,Rohrlich1965,Teitelboim1970,Teitelboim1971a,Teitelboim1971b,LandauLifshitz}. One typically starts by observing  how regularized
self-energy manifests as an acceleration term in the particle equation
of motion, which is easily recognized as a mass renormalization. This mass renormalization
is regularization-dependent, yet a finite subleading piece proves to reproduce
the ALD force precisely and universally.
Among these is an interesting treatment offered by Sidney Coleman~\cite{Coleman}, apparently a project he performed during
a summer internship, which treats the charge as truly point-like and regularizes
the Maxwell theory instead. We will review this in the next section, in part
because it offers the simplest derivation of the ALD force. 

Our aim here is a brute-force derivation of the self-force on a point charge undergoing an
arbitrary motion, when Maxwell is replaced by Born--Infeld. One might be
inclined to believe that nothing changes with Bremsstrahlung and perhaps with 
the associated self-force as well, since the asymptotic radiative field reduces
to the Maxwellian form. We will show  that this is not
the case and compute these physical quantities perturbatively. The very same
problem was in fact tackled by Feenberg \cite{Feenberg}, soon after the BI theory
was proposed. We shall review his approach later in an appendix, compare the
results against ours, and point out where he went astray; the reason behind the discrepancy should become clear
as we progress with the analysis.

In $d=4$, the  Born--Infeld action~\cite{BornInfeld} coupled to a point charge is\footnote{We record our conventions that include the
mostly-plus metric,
\bea
  \eta_{\mu\nu}=\operatorname{diag}(-1,+1,+1,+1)\ ,
  \qquad
  \tilde \cF^{\mu\nu}
  =\frac12\epsilon^{\mu\nu\rho\sigma}\cF_{\rho\sigma} \ ,
  \qquad
  \epsilon_{0123}=+1 \ . \nonumber
\eea
}
\bea
S_{\mathrm{BI}}+S_{\rm particle}=\int d^4x\,\cL_{\rm BI}-m_0\int d\tau+n\int\cA_\mu dz^\mu \ ,
\eea
with
\bea\label{ell_def}
 \cL_{\rm BI}=\frac{1}{\bbe^2\ell^4}(1-\mathscr T) \ , \qquad \mathscr T \equiv
  \sqrt{1+\frac{\ell^4 I}{2}-\frac{\ell^8 J^2}{16}} \ ,
\eea
whose two invariants are
\begin{equation}
  I=\cF_{\mu\nu}\cF^{\mu\nu} \ ,
  \qquad
  J=\cF_{\mu\nu}\tilde \cF^{\mu\nu} \ ,
\end{equation}
familiar from gauge theories. Note that the physical charge, in terms of which
physical expressions will be written in the end, is
\begin{equation}
  q=\bbe n \ ,
\end{equation}
coupled to the canonically normalized potential, $\cA_\mu/\bbe$.

The small $\ell$ expansion of $\mathscr T$ and the subsequent truncation to the first nontrivial order bring us back to the Maxwell theory with $-I/4\bbe^2$
as the Lagrangian density. In this sense, one could take the BI theory as a
regulated version of the Maxwell theory. On the other hand, the string-theory context would give us $\ell^2=2\pi\alpha'$, which is very much physical.
As advertised, we will take $\ell$ as a physical parameter and explore
its consequences. We will be limited to a perturbative analysis in the
end, unfortunately, tracing the nontrivial effect to the $\ell^3$ order
for quantities of interest here.

One place where the nonlinear nature of BI theory manifests most dramatically
is the Coulombic field of a stationary point charge. Let us start by writing
\begin{equation}\label{P}
  \cP^{\mu\nu}(\cF)
  \equiv -2\frac{\delta S_{\rm BI}}{\delta \cF_{\mu\nu}}
  =\frac{1}{\bbe^2\mathscr T}
  \left(
    \cF^{\mu\nu}-\frac{\ell^4 J}{4}\tilde \cF^{\mu\nu}
  \right) \ ,
\end{equation}
whereby the field equations are
\begin{equation} \label{BI_eq}
  \partial_\nu \cP^{\mu\nu}(\cF)=j^\mu\ ,
  \qquad
  \partial_{[\lambda}\cF_{\mu\nu]}=0 \ .
\end{equation}
This teaches us two things. First, the Gauss law and its resulting $1/r^2$ field due
to a point charge manifest with $\cP^{0i}$, the electric displacement $\mathbf D_i$.
Second, inverting the relation between $\cP$ and $\cF$,
we find that $\mathbf E_i=\cF_{i0}$ is entirely finite at the location of a point charge if not
regular.

When the charge is stationary, the exact solution in the rest frame $x=(t,\mathbf x)$
is,
\bea
\mathbf D=\frac{n}{4\pi r^2}\,\hat{\mathbf r} \ , \qquad \mathbf E=\frac{\bbe^2r^2}{\sqrt{r^4+r_{\rm BI}^4}}\,\mathbf D= \frac{\bbe^2n}{4\pi \sqrt{r^4+r_{\rm BI}^4}}\,\hat{\mathbf r} \ ,
\eea
where $\mathbf r=\mathbf x-\mathbf z_0$, $\mathbf z(\tau)=\mathbf z_0$, and
$r^2=\mathbf r^2$, while the magnetic counterparts vanish,
$\mathbf H=0=\mathbf B$. The length scale $r_{\rm BI}$ that regulates the solution is
such that
\bea\label{r_BI}
r_{\rm BI}^2 = \frac{\bbe^2|n|\ell^2}{4\pi} \ .
\eea
For more general motion, we boost both sides to a comoving frame
\bea
\cF_{\rm com}= \Sigma(r)\, \cP_{\rm com} \ ,
\eea
which will serve as the seed for solving the full field equation.
Here $\Sigma(r)$ is the comoving version of the above prefactor,
$\Sigma(r)=\bbe^2r^2/\sqrt{r^4+r_{\rm BI}^4}$, between
$\mathbf E$ and $\mathbf D$ of the stationary solution,
and again reduces the divergent behavior of $\cP_{\rm com}$ at $x= z(\tau)$ to a finite one.
The only difference from above is $\mathbf r=\mathbf x-\mathbf z(\tau)$, with $\mathbf x$ of the comoving frame.
By setting $\mathbf z(\tau)=0$ in the comoving frame, we have $r^2=\mathbf x^2$, which
we will use throughout this note.

When $z(\tau)$ remains inertial, the Bianchi identity holds
\bea
d\cF_{\rm com}\biggr\vert_{{\rm inertial}}=0 \ .
\eea
When the motion involves acceleration, $\cF_{\rm com}$ is no longer a solution, however.
Starting from the above expression under an arbitrary time-like trajectory $z(\tau)$,
we  must generally expand
\bea
\cF=\cF_{\rm com}+f \ ,
\eea
and solve the nonlinear BI equations for $f$ with $\cF_{\rm com}$ acting like an
additional source to $f$. Acceleration of $z(\tau)$ will feed into the corrective $f$,
eventually leading us to the desired expansion of Bremsstrahlung and the
self-force.

An astute reader familiar with how the self-force manifests for regulated
Maxwell theory may have caught something odd in this train of thought.
As one can see from Coleman's crystal clear derivation in the next section,
this force is nothing but the Lorentz force on the charge due to regulated
Maxwell fields generated by the charge itself, hence the expression ``self-force.''
In the Maxwell theory, it  starts with the acceleration
term, multiplied by divergent electromagnetic self-energy, followed
by a finite remainder that is regulator-independent and proportional to the time derivative
of acceleration. The latter is the radiation damping force, while the former renormalizes
the bare mass of the charged particle.

Trying to import the same thought here, however, fails miserably. $\cF_{\rm com}$
is already finite while its ill-defined direction-dependence washes out once we view the
charge as a distribution. The crucial subleading correction
$f$ due to acceleration cannot survive at $\mathbf x =\mathbf z(\tau)$.
In fact, we will later see that $f$ vanishes at least like $r$ at the origin
in the retarded solutions, and no Lorentz self-force due to acceleration results.
Of course this cannot be right; at least the finite self-energy
should be added to the effective mass of the charge, and should manifest as a term proportional to the acceleration in the particle equation
of motion. Since the latter has to involve $f$ somehow, despite the previous statement
of $f$ vanishing  at the origin, there must be a further subtlety with the
Born--Infeld theory despite its natural finiteness.

We will start with a lightning review of Coleman's self-force derivation and
move on to the Born--Infeld dynamics. A surprise in the latter, with $\cF$ naturally
finite at the point-charge location without regularization, is that the naive 
Lorentz force formula for a point charge would not catch the anticipated self-force. 
Although this is eventually resolved, the same also points us to a different and
more robust way of extracting the self-force from the on-shell action of the electromagnetic fields. 
We will start with a linear approximation in $f$, deriving
self-forces that are linear in the nonrelativistic acceleration $\mathbf a$ and its
time derivatives, and move on to a nonlinear completion that also serves as a
relativistic completion. The result is compared to an independent Bremsstrahlung
computation. Despite how the Born--Infeld theory
is arbitrarily close to Maxwell in the asymptotic region, the Bremsstrahlung
and the resulting dissipation of the particle kinetic energy receive
higher $\ell$ corrections. We close with a summary and three appendices.

\section{A Lookback on a Maxwellian Point Charge}

The problem of Maxwellian self-force has been addressed numerous times by renowned
physicists. One of the most complete treatments of the matter was by Teitelboim~\cite{Teitelboim1970,Teitelboim1971a,Teitelboim1971b},
which we will briefly revisit later, while a lesser known yet the most brilliant
treatise was offered by Sidney Coleman~\cite{Coleman}.
One should recall how the derivation of the self-force inevitably involves
a regularization that controls the divergence of the Coulombic self-energy.
This often takes the form of smearing of the point charge, yet such a smeared charge is at best cumbersome to handle.

Coleman's idea was to forget about such a smearing and instead replace
the Green's function that constructs the self-field solution by its
regulated version and push the computation of $\cF$. One
first solves for extrema of
\begin{equation}
  S_{\mathrm{Maxwell}}+S_{\rm particle}
  =-\frac{1}{4\bbe^2}\int d^4x \,\cF^2
  -m_0\int d\tau
  +n\int \cA_\mu(z(\tau))\,\frac{d z^\mu}{d\tau}d\tau \ ,
\end{equation}
to obtain $\cF$ for arbitrary motion $z(\tau)$, where in place
of the usual retarded Green's function, one uses the Pauli--Villars (PV)-regulated one\footnote{Since we would be using it merely for classical solutions, the usual
complaint about the Pauli--Villars regulator destroying the gauge redundancy is irrelevant.}
\bea\label{PV_A}
\cA^{(\Lambda)}_\mu(x)=\bbe^2n\int_{-\infty}^{\infty}d\tau\;  G^{(\Lambda)}_{\rm ret}(x-z(\tau))\, u_\mu \ .
\eea
The real space form of the PV-regulated Green's function is well known, if not often quoted,
\bea\label{ret}
G^{(\Lambda)}_{\rm ret}(x-z) & = & \vartheta(t-t_z) \vartheta(s^2)\frac{\Lambda}{4\pi s} J_1(\Lambda s) \ ,
\eea
where $\vartheta$ is the step function and $s^2\equiv -(x-z)^2= -\eta_{\mu\nu}(x-z)^\mu(x-z)^\nu$.
The particle's time coordinate is denoted as $t_z$.

The PV-regulated $\cF^{(\Lambda)}$ is an antisymmetrized partial derivative of this $\cA^{(\Lambda)}_\mu$,
\bea
\cF^{(\Lambda)}_{\mu\nu}(x)=2\bbe^2n\int_{-\infty}^{\infty}d\tau\;  \partial^x_{[\mu} G^{(\Lambda)}_{\rm ret}(x-z(\tau))\, u_{\nu]} \ .
\eea
For the self-force
problem, we need $z(\tau_0)=0$ for some $\tau_0$. We will also set $\tau_0=0$.
Even though the field is finite everywhere, $\cF^{(\Lambda)}$  actually suffers
from a directional ambiguity at the point-charge position. This ambiguity
can be easily removed for the Lorentz self-force by considering the charge as a ``distribution,'' which is the same
as smearing it to a spherically symmetric configuration to be taken to be point-like at the end of the computation.
This is effectively the same as dropping the derivative on the step function in $\vartheta(s^2)$
and also the same as taking an angular average of the field strength,
$\langle\cF_{\mu\nu}^{\rm self}(0)\rangle^{\rm angular}$.

Let us adopt a simplified notation $\vartheta(t-t_z) \Delta (s)\equiv G^{(\Lambda)}_{\rm ret}(x-z(\tau))$
and subsequently work out $\cF^{(\Lambda)}$ at $x=0$, where $s^2=-z(\tau)\cdot z(\tau)$.
With the chain rule,
\bea
\frac{\partial \Delta}{\partial x^\mu}=\frac{z_\mu}{s}\frac{\partial \Delta}{\partial s} \ ,
\eea
and $u_\mu d\tau=dz_\mu$,  we find
\bea
\langle\cF_{\mu\nu}^{\rm self}(0)\rangle^{\rm angular}&=& -\bbe^2n\int_{0}^{\infty}\left( \frac{z_\mu}{s} \frac{d\Delta(s)}{ds}\frac{dz_\nu}{ds} - \frac{z_\nu}{s} \frac{d\Delta(s)}{ds}\frac{dz_\mu}{ds}\right) ds \cr\cr
& = & \bbe^2n\int_{0}^{\infty}\Delta(s)\frac{d}{ds}\left( \frac{z_\mu}{s}\frac{dz_\nu}{ds} - \frac{z_\nu}{s} \frac{dz_\mu}{ds}\right) ds \ .
\eea
The potential boundary contribution at $s=0$ in this integration by parts vanishes.

The Taylor expansion in $s$ of the quantity inside the parentheses leads us to
the desired $\cF_{\mu\nu}^{\rm self}(0)$ and the Lorentz self-force $n \cF_{\mu\nu}^{\rm self}(0)u^\nu$.
The particle passes through $x=0$ at $\tau=0$, so the coefficients
of the Taylor expansion in $s$, constructed from derivatives of $z$ at $s=0$, can be
translated to the latter's derivative with respect to $\tau$. For instance, one finds
\bea
\frac{d z_\mu}{ds}\biggr\vert_{s=0} =-u_\mu\biggr\vert_{\tau=0} \ ,
\eea
where one should take note of the sign. Such relations occur because $\tau$ is
meant to be, by definition, the proper time defined by the worldline itself
rather than some random parameter; for the past trajectory very
near $x$,  $\tau\simeq -s$ is inevitable once we use the harmless additive shift
of $\tau$ to set $z(0)=x$.  For the self-force computation  at hand, we also need
\bea
\frac{d^2 z_\mu}{ds^2}\biggr\vert_{s=0}=a_\mu\biggr\vert_{\tau=0}\ ,\qquad
\frac{d^3 z_\mu}{ds^3}\biggr\vert_{s=0}=\left(-\dot a_\mu+\frac14a^2u_\mu\right)\biggr\vert_{\tau=0} \ ,
\eea
where the dot denotes the derivative with respect to the proper time. These allow us to express $\cF^{(\Lambda)}_{\mu\nu}(0)$ in terms of kinematic
variables of the charged particle at the point as its worldline passes that very same point.
The relevant coefficients are moment integrals of $\Delta$, of which only the following two,
\bea
\int_0^\infty \Delta(s)\,d s
=\frac{\Lambda}{4\pi}\ ,\qquad
\int_0^\infty s\Delta(s)\, d s
=\frac{1}{4\pi} \ ,
\eea
survive the limit $\Lambda\rightarrow \infty$.

Recall how the integrals extend to arbitrarily large $s>0$, or equivalently to
arbitrary past of the particle trajectory. This comes about because the
regulated Green's function is supported inside the light-cone rather than on
the light-cone. Nevertheless, we end up with the first few moment integrations
dominating the answer because in effect the actual contribution to (\ref{PV_A}) is
tightly concentrated on the past light-cone. With our Pauli--Villars regulator,
this happens thanks to the highly oscillatory behavior of $J_1(\Lambda s)$ away from $s=0$;
the off-light-cone region, say, $s>1/\Lambda$, contributes to $\cF^{(\Lambda)}$ by
a very small amount suppressed by inverse powers of $\Lambda$,
which in turn justifies the above Taylor expansion Coleman employed.

Evaluating the resulting Lorentz self-force on the particle of charge $n$,  we find
\begin{equation}
m_0 a^\mu =F^\mu_{\rm ext}+ F_{\rm self}^\mu\ , \qquad
F_{\mathrm{self}}^\mu
=-\frac{q^2 \Lambda}{8\pi}a^\mu
+\frac{q^2}{6\pi}
\left(\dot a^\mu-a^2u^\mu\right)
+\order(\Lambda^{-1}) \ ,
\label{self-force-final}
\end{equation}
with $q=\bbe n$.
The first term of $F_{\rm self}^\mu$ is moved to the left-hand side of Newton's law and renormalizes the bare mass $m_0\rightarrow m_0+q^2\Lambda/8\pi$, while the
second is the celebrated ALD force. Since $\Lambda$ is introduced
as a regulator rather than as a physical parameter, we customarily take
$\Lambda\rightarrow\infty$ which removes the subleading terms.
The resulting dissipative ALD force accounts for how
the particle kinetic energy is drained and converted to the radiation power.
Different regulators may result in different mass renormalization but the finite
ALD force itself is universal.

A couple of customary observations regarding this ALD damping force are in order.
One is that the resulting particle equation of motion admits unphysical runaway solutions, due to
the new 3rd-order time derivative of $z$. In more general
electrodynamics, such as in the Born--Infeld theory next, this becomes potentially worse and worse
since higher time derivatives of $a$ could enter the story. Although one often encounters
discussion about this as a puzzle, this is hardly one. It is merely an artifact of the
perturbative treatment, which one can safely ignore in a context with proper initial/final conditions which are part of physics.

The other is how the uniform
acceleration, $\dot a^\mu =a^2u^\mu$, turns off the self-force entirely while the radiation power continues on. There is no
issue with energy conservation thanks to how there are three separate components of energy involved:
particle kinetic energy combined with inflow of energy due to the external motion-driving force,
the near-field Coulombic energy, and the radiation energy. Related to this is how
the self-force may be categorized into two types, dissipative and reversible.
The latter is also known as the Schott type and acts as an intermediary among the three components of energy, while the dissipative term connects the particle energy and the radiation
power more directly. For motions that start and end inertially, this energy accounting
is quite straightforward. Decades ago, Boulware~\cite{Boulware} cleared up additional conceptual puzzles that
seemingly appear for a perennial uniform acceleration. We highly recommend this work
to those new to this phenomenon.

\section{Born--Infeld Field in Comoving Frame}
\label{sec:tube-setup}

For the computation, it is simplest to work in the comoving frame, which is equivalent to
labeling spacetime points as
\begin{equation}
z(\tau)+{\mathbf x}^i\,e_i(\tau) \ ,
\end{equation}
where the $e_i$ form an orthonormal triad sitting at the particle position, ${\mathbf x}=0$, obeying
\bea
\frac{de_i}{d\tau}=(a\cdot e_i)\,u\equiv \mathbf a_i(\tau)\,u \ ,
\eea
with  the particle's 4-velocity $u(\tau)$ and acceleration $a(\tau)$. Its spatial part in this comoving frame is written as $\mathbf a$, henceforth.

Taking the time coordinate $t$ for the span of $\mathbf x$
to be the same as the proper time $\tau$ of the charge, small displacements of spacetime position can
then be written as a set of 1-forms,
\begin{equation}
u\,dt+{\mathbf x}^i\,\frac{de_i}{dt}\,dt+e_i\,d{\mathbf x}^i=(1+{\mathbf a}\cdot{\mathbf x})\,u\,dt+e_i\,d{\mathbf x}^i \ ,
\end{equation}
whereby the Minkowski metric becomes
\begin{equation}
ds^2=-\cN^2dt^2+\delta_{ij}\,d{\mathbf x}^id{\mathbf x}^j \ ,\qquad \cN(t,{\mathbf x})=1+{\mathbf a}(t)\cdot{\mathbf x} \ .
\label{metric_com}
\end{equation}
The spatial distance $r$ from the point charge is such that $r^2=\mathbf x^2$ in this comoving frame,
since the point charge sits at $\mathbf x=0$.

We set $c=1$ and work with electric and magnetic components, defined
in this frame  as
\begin{eqnarray}
\mathcal F &=& {\mathbf E}_i\,d{\mathbf x}^i\wedge(\cN\,dt)+\star_3{\mathbf B}\ , \cr\cr
\tilde{\mathcal P} &=& -{\mathbf H}_i\,d{\mathbf x}^i\wedge(\cN\,dt)+\star_3{\mathbf D} \ .
\end{eqnarray}
Here $\star_3\mathbf v=\epsilon_{ijk}\mathbf v_i\,d\mathbf x^j\wedge d\mathbf x^k/2$
with $\epsilon_{123}=1$, and $\tilde \cP$ is the Hodge dual of
$\cP$. These are
related among themselves as
\bea
{\mathbf E}=\frac{\partial\mathcal H}{\partial{\mathbf D}}\ ,\qquad {\mathbf H}=\frac{\partial\mathcal H}{\partial{\mathbf B}} \ ,
\eea
via the BI energy density
\bea
\mathcal H=\frac{1}{\bbe^2\ell^4}\Big(\sqrt{1+\bbe^4\ell^4{\mathbf D}^2+\ell^4{\mathbf B}^2+\bbe^4\ell^8({\mathbf D}\times{\mathbf B})^2}-1\Big) \ .
\eea
More concretely, we have
\bea
 \mathbf{E}&=&\bbe^2\frac{(1+\ell^4\mathbf{B}^2)\mathbf{D}-\ell^4(\mathbf{D}\cdot\mathbf{B})\mathbf{B}}
 {\sqrt{1+\bbe^4\ell^4\mathbf{D}^2+\ell^4\mathbf{B}^2+\bbe^4\ell^8|\mathbf{D}\times\mathbf{B}|^2}}\ ,\cr\cr
 \mathbf{H}&=&\frac1{\bbe^2}\frac{(1+\bbe^4\ell^4\mathbf{D}^2)\mathbf{B}-\bbe^4\ell^4(\mathbf{D}\cdot\mathbf{B})\mathbf{D}}
 {\sqrt{1+\bbe^4\ell^4\mathbf{D}^2+\ell^4\mathbf{B}^2+\bbe^4\ell^8|\mathbf{D}\times\mathbf{B}|^2}} \ ,
 \label{tube-constitutive}
\eea
sometimes referred to as ``constitutive'' relations.

Here we set up a perturbative scheme that solves the field equations (\ref{BI_eq})
\bea
d\mathcal F=0\ ,\qquad d\tilde{\mathcal P}=J \ ,
\eea
where $J=n\,\delta^{(3)}({\mathbf x})\,d^3{\mathbf x}$ is a 3-form current
for a point charge undergoing an arbitrary motion.
With $\nabla_\mu$ the spacetime covariant derivative, the latter is also written as
\bea
 \nabla_\nu\cP^{\mu\nu}(\cF)=j^\mu \ ,
\eea
with the vectorial current $j$. Outside the worldline, the second field equation reads $d\tilde{\cP}=0$.
Component-wise in the comoving frame, we would solve
\begin{equation}
 \begin{aligned}
\nabla_{\mathbf x}\cdot\mathbf{B}&=0\ ,&
- \partial_t\mathbf{B}&=\nabla_{\mathbf x}\times[\cN\mathbf{E}]\ ,\\
 \nabla_{\mathbf x}\cdot\mathbf{D}&=n\delta^{(3)}(\mathbf x)\ ,\quad&
  \partial_t\mathbf{D}&=\nabla_{\mathbf x}\times[\cN\mathbf{H}] \ ,
 \end{aligned}
 \label{tube-field-equations}
\end{equation}
with the lapse function $\cN=1+\mathbf{a}\cdot\mathbf x$. Without loss of generality, 
we will take $n>0$ for the rest of the note. 

We will expand the field equations for $\cF$ around $\cF_{\rm com}$
and their counterparts for $\cP(\cF)$, and solve them perturbatively
with the BI length scale $\ell$ of (\ref{ell_def}) as the expansion parameter.
With our point charge at $\mathbf x=0$, we start from
\bea
\mathcal F_{\rm com}=\frac{dr\wedge(\cN\,dt)}{\ell^2\sqrt{1+y^4}}\ ,\qquad \tilde{\mathcal P}_{\rm com}=\frac{n}{4\pi}\,d\Omega\ , \qquad y\equiv \frac{r}{r_{\rm BI}} \ ,
\eea
which would be a solution if and only if the motion is inertial.
Here the fundamental length scale $r_{\rm BI}$ of the point charge,
defined in (\ref{r_BI}), is the radius at which the Coulomb field begins to reach
the maximal BI value $\ell^{-2}$.

More precisely, we expand as\footnote{The coefficients with $N>0$ have dimension ${\rm length}^{-N}$.
We count $\mathbf{a}^{(N)}=d^N\mathbf{a}/d\tau^N$ with weight $N+1$ and retain,
at each field order, all motion products whose total weight equals that order.}
\begin{equation}
 \begin{aligned}
 \mathbf{E}&=\ell^{-2}\sum_{N=0}^{\infty}r_{\rm BI}^N\mathbf{e}_N\ ,&\quad
 \mathbf{D}&=\frac1{\bbe^2\ell^2}\sum_{N=0}^{\infty}r_{\rm BI}^N\mathbf{d}_N\ ,\\
 \mathbf{B}&=\ell^{-2}\sum_{N=0}^{\infty}r_{\rm BI}^N\mathbf{b}_N\ ,&\quad
 \mathbf{H}&=\frac1{\bbe^2\ell^2}\sum_{N=0}^{\infty}r_{\rm BI}^N\mathbf{h}_N \ .
 \end{aligned}
 \label{tube-field-series}
\end{equation}
The zeroth order is the exact static point-charge field corresponding to $\cF_{\rm com}$, or
equivalently,
\begin{equation}
 \mathbf{d}_0=\frac{\hat{\mathbf{r}}}{y^2} \ ,\qquad
 \mathbf{e}_0=\frac{\hat{\mathbf{r}}}{\sqrt{1+y^4}} \ ,\qquad
 \mathbf{b}_0=\mathbf{h}_0=0 \ ,
 \label{tube-static-seed}
\end{equation}
where we took out a normalization factor. Restoring the latter gives
\bea
 \mathbf{D}_{\rm com}=\frac{\mathbf{d}_0}{\bbe^2\ell^2}
 =\frac{\hat{\mathbf{r}}}{\bbe^2\ell^2y^2}
 =\frac{n}{4\pi r^2}\hat{\mathbf{r}} \ ,
\eea
with its finite electrostatic energy
\bea
M_{\rm BI}=\eta \frac{q^2}{6\pi r_{\rm BI}}\ , \qquad
 \eta=\int_0^\infty\frac{dy}{\sqrt{1+y^4}}
 =\frac{\Gamma(1/4)^2}{4\sqrt\pi} \ .
 \label{BI_mass}
\eea

Subtracting this comoving seed from the full two-forms gives
\begin{equation}
\begin{aligned}
 f&=\frac1{\ell^2}\sum_{N\ge1}r_{\rm BI}^N
 \left[\mathbf e_{N i}\,d\mathbf x^i\wedge
 (\cN\,dt)+\star_3\mathbf b_N\right] \ ,\\
 \tilde p&=\frac1{\bbe^2\ell^2}\sum_{N\ge1}r_{\rm BI}^N
 \left[-\mathbf h_{N i}\,d\mathbf x^i\wedge
 (\cN\,dt)+\star_3\mathbf d_N\right] \ .
\end{aligned}
\label{tube-correction-components}
\end{equation}
In other words, we perform a perturbative expansion of $\cF$
\bea
\cF=\cF_{\rm com}+f\ , \qquad f=f_1+f_2+f_3+\cdots \ ,
\eea
which results in an iterative set of equations for $f_{N}$'s to be
solved order by order in  $r_{\rm BI}$.
The Euler--Lagrange half of the equations $d\tilde{\mathcal P}=J $ reduces to
\bea
d\tilde p=0 \ ,
\eea
with the same $\tilde p$ above, or equivalently with $p\equiv \cP(\cF)-\cP(\cF_{\rm com})$. Although this one looks deceptively simple,
it is in fact the more involved of the two halves since we need to solve $p$ as an expression
of $f$ and $\cF_{\rm com}$ which enter the definition of $\cP$ and $p$ nonlinearly.
The other half is of course the Bianchi identity,
\bea
df=-d\cF_{\rm com} \ ,
\eea
where the right-hand side is linear in the instantaneous acceleration $a(t)$; 
its time derivatives $a^{(k)}$, weighted by powers of $r_{\rm BI}$, enter through the iteration of the field equations.

In the next section, we solve $f_N$'s iteratively and feed them to what we will later call the tubular self-force
that involves both $\cP$ and $\cF$. The resulting self-force can be organized into the powers of $\ell$ as
\bea\label{self_force_covariant_scheme}
F_{\rm self}\sim \#\frac{1}{\ell}a + \#(\dot a-a^2u) + \sum_{N=1} \#\ell^N (a^{(N+1)} +\cdots) \ .
\eea
Each summand can be further expanded into powers of $a$ and $a^{(k)}$'s, as shown in the $\ell^0$ term,
whose form is on par with the ALD term of the Maxwell theory, appropriately weighted by powers of $1/c^2$, which we suppress here. Here we see that, if we take the attitude of taking $1/\ell$ as a regulator to
Maxwell, we find the same usual answer to the self-force in the $\ell\rightarrow 0$ limit; the first term renormalizes the mass by self-energy $\sim 1/\ell$ while the second is nothing but the ALD force.

With finite $\ell$, we shall actually arrive at such a relativistic expression through
nonrelativistic computations in the comoving frame that would give a double expansion,
\bea\label{self_force_NR_scheme}
\mathbf F_{\rm self}\sim \#\frac{1}{\ell}\mathbf a + \#(\dot {\mathbf a} + \cdots) + \sum_{N=1} \#\ell^N (\mathbf a^{(N+1)} +\cdots) \ ,
\eea
where the leading $\mathbf a^{(k)}$ contribution in each summand arises from $f_{k+1}$. At each nonnegative power
$\ell^N$, $N\ge 0$, we find a sequence of contributions from  $f_{N+2}$,  $f_{N+1}f_1$, $f_{N}f_2$, \dots, $f_{N}f_1f_1$,  $f_{N-1}f_2f_1$, etc., generating a sequence of nonlinear combinations of lower time derivatives of
${\mathbf a}$.\footnote{In certain computations for the Maxwell theory, say, \`{a} la Teitelboim,
the same type of power counting works, although the linearity allows only terms up to quadratic order in $f$; $f_1$ for the mass renormalization term and $f_2$ and $f_1f_1$ for the ALD term.}
Although the computations are performed in the comoving frame, these nonlinear terms sum up
to a fully relativistic expression of the kind (\ref{self_force_covariant_scheme}) without us
having to ``complete'' the self-force by adding additional pieces at higher order of
$1/c^2$'s.

\section{Tubular Self-Force}
\label{sec:tubular}

Following the variational principle with the combined action,
we find the usual Lorentz force equation from the variation of
\begin{equation}
S_{\rm particle}
  =
  -m_0\int d\tau +S_{\rm minimal}\ ,\qquad S_{\rm minimal}=n\int \cA_\mu(z(\tau))\,\frac{d z^\mu}{d\tau}d\tau \ ,
\end{equation}
which gives, after solving for $\cA_{\rm self}$ due to the same particle,\footnote{As in Coleman's
treatise, the angular average $\langle\cF_{\rm self}^{\mu\nu}\rangle^{\rm angular}$ emerges naturally
once we view the point charge as a distribution. }
\bea\label{self_L}
m_0 a^\mu= F^\mu_{\rm ext}+ n\langle\cF_{\rm self}^{\mu\nu}\rangle^{\rm angular}u_\nu \ .
\eea
We allow an external
force $F^\mu_{\rm ext}$ that drives the motion.
The notion of self-force emerges usually from this because  the regulated
field $\cF_{\rm self}$ knows about the
motion of the charge itself.

Previously we followed Coleman and saw how the Lorentz self-force, the second piece on the right,
accounts for the mass renormalization due to self-energy and the required dissipation,
for the example of classical Maxwell electrodynamics. With the PV-regulated Green's function, a divergent part of the self-force proportional to $a^\mu$ can be moved
to the left-hand side and renormalizes the bare mass $m_0$, while the remaining finite part emerges as the ALD radiation damping term.

In the Born--Infeld theory with $\cF_{\rm self}$ finite everywhere, on
the other hand, the above form of Lorentz self-force vanishes identically
regardless of detailed particle motion, as can be seen from the general perturbative scheme. 
The self-force would have to come from the subleading $f$ generated by accelerated motion,
as in $\cF_{\rm self}=\cF_{\rm com}+f$, since upon the angular average
the ambiguous  $\cF_{\rm com}$ at the origin cancels out. However, this $f$
also vanishes quickly at $x=z(\tau)$ and thus no self-force appears to be
present if we use (\ref{self_L}), as long as the solution due to the point source
can be trusted. In Section~\ref{sec:lorentz-tubular}, we will revisit this statement and discover that
something more subtle is going on, but for now let us trust this naive result at
face value.

Then, the only remaining logical possibility is to include the variation of the
on-shell expression of $S_{\rm BI}$. One can see
immediately that it has to be the case by considering the self-energy. The finite self-energy
implies that, for a slowly moving point charge, the on-shell action of the field reduces to\footnote{We took 
the gauge choice $\cA(\infty)=0$ for the evaluation of $S_{\rm minimal}^{\rm self}$, which 
does not enter the self-force computation down the road.}
\bea\label{self_action}
S^{\rm self}= S_{\rm BI}^{\rm self}+S_{\rm minimal}^{\rm self}\equiv (S_{\rm BI}+S_{\rm minimal})\biggr\vert_{\cF=\cF_{\rm self}} \simeq -M_{\rm BI} \int d\tau\ .
\eea
Here,
\bea 
M_{\rm BI}\equiv \eta\,\frac{q^2}{6\pi r_{\rm BI}} 
\eea
is precisely the self-energy due to the BI field generated by the point charge, 
so that (\ref{self_action}) represents the correct mass renormalization of the
charged particle in question, $m_0\rightarrow m_0+ M_{\rm BI}$.

Therefore, the variation of $S^{\rm self}$ with respect to the particle position
$z^\mu$ has to yield a self-force term of the type
\bea
F^\mu_{\rm self}= -M_{\rm BI} a^\mu +\cdots \ ,
\eea
at least. The lesson from this is that we must consider the variation under
$z\rightarrow z+\xi$ of the on-shell version $S_{\rm BI}^{\rm self}$ of the action
$S_{\rm BI}$ as well to complete the particle equation of motion. To carry out the computation, let us split the
self-action into two parts,
\bea
S^{\rm self} = S^{\rm self}_{\rm out} + S^{\rm self}_{\rm in} \ .
\eea
$S^{\rm self}_{\rm out}$  is computed by restricting the integration region for
$S_{\rm BI}^{\rm self}$ to $r>\epsilon$ while the remainder  from $r<\epsilon$ 
combined with $S_{\rm minimal}^{\rm self}$ defines $ S^{\rm self}_{\rm in} $.
The artificial boundary that separates the two regions will be denoted as
$\mathbb S^2_\epsilon(z)\ltimes\mathbb R$.

Let us work out what happens if we vary $z(\tau)$ inside $S^{\rm self}_{\rm out}$.
The variation of $\cL_{\rm BI}^{\rm self} $
with respect to small $\delta z=\xi$ shifts the integrand of the BI action
on shell,
\bea\label{PF_variation}
\cV \,\delta_\xi\cL_{\rm BI}^{\rm self}=-\tilde \cP_{\rm self}\wedge \delta_\xi(\cF_{\rm self})
= \tilde \cP_{\rm self} \wedge d(\xi\lrcorner \cF_{\rm self}) \ ,
\eea
when written via the volume form $\cV$. Here $\delta_\xi$ is meant to act
only on $\cA$ and $\cF$, and not on the metric, since it is really from the
position variation of the charged particle. The sign flip in the last step
is due to the same; we varied  $z$ by $+\xi$ but the on-shell
field $\cF$ depends on the combination $x-z(\tau)$, so that
$\delta z=\xi$ translates to a Lie derivative $\cL_{-\xi}$ with respect to $x$.

Upon an integration by parts, the bulk part $\sim d\tilde \cP\wedge (\xi\lrcorner\cF)$ of
(\ref{PF_variation}) vanishes by the field equation, with no charge outside
the tubular region assumed, so at most a boundary contribution survives.
This  boundary term from the above variation of $\cL_{\rm BI}^{\rm self}$
then produces
\bea\label{shift1}
\lim_{\epsilon\rightarrow 0}  \int_{\mathbb S^2_\epsilon(z)\ltimes\mathbb R}\tilde\cP_{\rm self} \wedge (\xi\lrcorner \cF_{\rm self}) \ ,
\eea
with $\mathbb R$ the time direction. For the moment, the boundary orientation
remains such that the normal vector points inward towards the worldline.
$\cF_{\rm self}$ is finite at $z$ while $\cP_{\rm self}\sim 1/r^2$ typically; the two can combine to produce a finite contribution here.

Although (\ref{shift1}) will prove enough for the eventual self-force
computation, it is important to realize that  the domain of
integration itself, defined as ``outside'' of the infinitesimal tubular neighborhood
around the particle position, also shifts under $z\rightarrow z+\xi$.
This shift of $\mathbb S^2_\epsilon(z)$ generates a bulk integral of the Lagrangian density
in a puffed-up cylinder bounded by  $\mathbb S^2_\epsilon(z+\xi)\ltimes\mathbb R - \mathbb S^2_\epsilon(z)\ltimes\mathbb R$.
This additional variation of $S_{\rm out}^{\rm self}$ is then
\bea\label{shift2}
\lim_{\epsilon\rightarrow 0}  \int_{\mathbb S^2_\epsilon(z)\ltimes\mathbb R}\xi\lrcorner \cV \cL_{\rm BI}^{\rm self} \ .
\eea
Note that, here, the displacement of $z$ by $\xi$ shifts the cylinder position as $x\rightarrow x+\xi$,
which is opposite to how $x$ shifted under the above variation that led to (\ref{PF_variation}).

Together, (\ref{shift1}) and (\ref{shift2}) produce a self-force term from variation of  $S^{\rm self}_{\rm out}$
in the particle equation of motion as
\bea\label{shift}
\delta S^{\rm self}_{\rm out} \quad\Rightarrow \quad 
F_{\rm tubular}^\mu=- \lim_{\epsilon\rightarrow 0}  \int_{\mathbb S^2_\epsilon(z)}(T_{\rm self})^{\mu}_{\;\;\alpha} {\mathbf n}^\alpha \ ,
\eea
where $(T_{\rm self})^{\mu}_{\;\;\alpha} $ is
\bea
T^\mu_{\;\;\alpha} =\cP^{\mu\beta}\cF_{\alpha\beta}+\delta^\mu_{\;\;\alpha}\cL_{\rm BI} \ ,
\eea
evaluated on the on-shell field $\cF_{\rm self}$. The minus sign comes from how the natural
orientation of the tubular boundary is inward in the above computation, yet
here  we use the outward normal
$\mathbf n$ to fit the common convention. In the instantaneous rest frame,
$\mathbf n =(0,\hat{\mathbf r})$ on $\mathbb S^2_\epsilon(z)$. 
We call this  effect in  (\ref{shift}) the ``tubular self-force.''

For the rest of the section, we will evaluate this self-force $F_{\rm tubular}$ and find that
it supplies the anticipated answers in the leading $\ell$ expansion, namely both
the correct mass renormalization of $m_0\rightarrow m_0+M_{\rm BI}$ and the subsequent 
$\ell$-independent ALD force. Since $\ell$ is not a regulator but a physical parameter of
this theory, subleading contributions in positive powers 
of $\ell$ are also present and must be considered physical. 

The self-force $F_{\rm tubular}$ comes from variation of $S^{\rm self}_{\rm out}$ only, albeit in the
limit $\epsilon\rightarrow 0$,  so a further thought on what $S^{\rm self}_{\rm in}$ may do
is needed. Given how this $F_{\rm tubular}$ produces sensible answers on its own, one could 
imagine  that no further contribution arises from $S^{\rm self}_{\rm in}$, yet this sounds 
very odd; as noted already, the naive Lorentz self-force part of the latter vanishes 
while the boundary contribution associated with the $S^{\rm self}_{\rm BI}$ part of 
$S^{\rm self}_{\rm in}$ should be merely the opposite of that of $S^{\rm self}_{\rm out}$. 
We will devote a later section to this question, where we take a deeper look
at what ``point'' charge really means, with the end result that $F_{\rm tubular}$ 
in (\ref{shift}) is indeed the entire self-force.

\subsection{Linear Contributions in Comoving Frame}
\label{sec:tube-linear-calculation}

As noted in the previous section, self-force terms weighted by 
$r_{\rm BI}^{N-2}\sim r_{\rm BI}^N/\ell^2$  
come from $f_{N}$ at the linear level and from combinations of
$f_k$'s with the sum of $k$ equal to $N$ at the nonlinear level. For convenience, 
we will refer to this set of contributions collectively as ``field order'' $N$.

Let us start by stating boundary conditions, both at the core and
in the asymptotic region. At each retained order $N\ge1$, we impose the core conditions
\begin{equation}
 \mathbf{d}_N=O(y^{-2}) \ ,\qquad \mathbf{b}_N=O(y) \ .
 \label{tube-core-conditions}
\end{equation}
We assume differentiable radial asymptotic expansions with a finite
limit of $y^2\mathbf{d}_N$. These expansions and bounds are assumed uniform
in angle and on compact proper-time intervals, including the retained
time derivatives.
The leading $y^{-2}$ displacement is then radial by the constitutive
relation and Faraday's law, as derived in \eqref{tube-core-radial-consequence}.

The exterior solution obeys the retarded condition with no independent
incident field. Let $a_{\max}$ be an acceleration bound and $T$ a variation
time scale, with $|\mathbf a^{(N)}|\lesssim a_{\max}/T^N$ for the retained
derivatives. The inner and weak-field exterior expansions overlap in
\bea
 r_{\rm BI}\ll r\ll\min\{a_{\max}^{-1},T\} \ .
\eea
In this overlap, the large-$y$ inner fields, re-expanded using
$y=r/r_{\rm BI}$, must match the retarded exterior fields term by term
at each power of $r_{\rm BI}$ as in \eqref{tube-inner-outer-powers}.
This fixes the remaining inner constants and outgoing multipole amplitudes.

For fields satisfying the conditions above, only the radial electric
stress survives the shrinking-tube integral in \eqref{shift}, giving
\begin{equation}
 \mathbf{F}_{\rm tubular}\big|_{N}
 =\frac{q^2r_{\rm BI}^{N-2}}{16\pi^2}
 \lim_{y\to0}y^2\int\hat{\mathbf{r}}\,(\mathbf{d}_N\cdot\hat{\mathbf{r}})\bigr\vert_{y}\,d\Omega \ .
 \label{tube-core-trace}
\end{equation}
Here $d\Omega$ is solid angle. The energy flux through the shrinking sphere
vanishes as $O(r)$, confirming $F_{\rm tubular}^0=0$ in the instantaneous rest frame. Field order $N$ contributes at force
order $r_{\rm BI}^{N-2}$ through the $l=1$ radial displacement coefficient.

With the inertial seed \eqref{tube-static-seed}, indicated by the subscript `0' below,
the linear response matrices $\mathsf M,\mathsf K$ have components
\begin{equation}
 \begin{aligned}
 (\mathsf M)_{ij}&\equiv\left.\bbe^{-2}\frac{\partial \mathbf E_i}{\partial \mathbf D_j}\right|_0
 =\frac{y^6\hat r_i\hat r_j}{(1+y^4)^{3/2}}
 +\frac{y^2(\delta_{ij}-\hat r_i\hat r_j)}{\sqrt{1+y^4}} \ ,\\
 (\mathsf K)_{ij}&\equiv\left.\bbe^2\frac{\partial\mathbf H_i}{\partial\mathbf B_j}\right|_0
 =\frac{y^2\hat r_i\hat r_j}{\sqrt{1+y^4}}
 +\frac{\sqrt{1+y^4}}{y^2}(\delta_{ij}-\hat r_i\hat r_j) \ .
 \end{aligned}
 \label{tube-jacobians}
\end{equation}
At this zeroth order, the other derivatives such as $\partial\mathbf H /\partial \mathbf D$ vanish.
At $N\ge1$, the constitutive expansion separates the unknown fields
from the nonlinear remainders $\bm{\mathcal E}_N,\bm{\mathcal H}_N$,
\begin{equation}
 \mathbf{e}_N=\mathsf M\mathbf{d}_N+\bm{\mathcal E}_N \ ,\qquad
 \mathbf{h}_N=\mathsf K\mathbf{b}_N+\bm{\mathcal H}_N \ .
 \label{tube-constitutive-order}
\end{equation}
Substituting the fields through order $N-1$ into \eqref{tube-constitutive}
gives the remainders as the $r_{\rm BI}^N$ coefficients of $\ell^2\mathbf{E}$
and $\bbe^2\ell^2\mathbf{H}$, respectively. Each nonlinear product contains at least
two positive-order fields whose orders sum to $N$. Thus every positive-order factor has
order below $N$, and $\bm{\mathcal E}_{1}=\bm{\mathcal H}_{1}=0$.

The constitutive expansion above relates the components of $f$ and
$\tilde p$, with $\mathbf d_N,\mathbf b_N$ as the unknown fields.
Since the seed satisfies $\nabla_\nu\mathcal P_{\rm com}^{\mu\nu}=j^\mu$,
the equations for the corrections are
\bea
 df=-d\mathcal F_{\rm com} \ ,\qquad d\tilde p=0 \ .
\eea
Their component form is obtained by subtracting the seed from
\eqref{tube-field-equations}.
We substitute \eqref{tube-field-series} and \eqref{tube-constitutive-order}
into \eqref{tube-field-equations} and use
$\nabla_{\mathbf x}=r_{\rm BI}^{-1}\nabla_y$. Multiplying by $r_{\rm BI}$ and
extracting the coefficient at order $N\ge1$ gives
\begin{equation}
 \begin{aligned}
 \nabla_y\times(\mathsf M\mathbf{d}_N)
   &=-\partial_t\mathbf{b}_{N-1}
     -\nabla_y\times[\bm{\mathcal E}_{N}+(\mathbf{a}\cdot\mathbf{y})\mathbf{e}_{N-1}] \ ,\\
 \nabla_y\times(\mathsf K\mathbf{b}_N)
   &=\partial_t\mathbf{d}_{N-1}
     -\nabla_y\times[\bm{\mathcal H}_{N}+(\mathbf{a}\cdot\mathbf{y})\mathbf{h}_{N-1}] \ ,\\
 \nabla_y\cdot\mathbf{d}_N&=\nabla_y\cdot\mathbf{b}_N=0 \ .
 \end{aligned}
 \label{tube-full-recursion}
\end{equation}
The charge is carried entirely by
$\nabla_y\cdot\mathbf{d}_0=4\pi\delta^{(3)}(\mathbf{y})$.
Projection onto spherical harmonics of angular degree $l$ gives radial
equations for $\mathbf{d}_N,\mathbf{b}_N$, with sources fixed by lower orders. After solving for these fields,
\eqref{tube-constitutive-order} supplies $\mathbf{e}_N,\mathbf{h}_N$ needed at order $N+1$.

The boundary conditions stated above were taken to apply to both linear
and nonlinear  terms, separately. In the local polynomial expansion for
general motion,  terms with different total degrees in $\mathbf a$ and its derivatives
match separately. At each degree, the conditions are imposed on the
sum of the particular and homogeneous fields.

At each length order, terms linear in $\mathbf{a}$ and its proper-time
derivatives are sourced only by the static seed and lower-order linear
fields in \eqref{tube-full-recursion}.
Rotational symmetry and parity restrict the electric response linear in
$\mathbf{a}^{(N)}$ to its radial and tangential projections.
The divergence-free condition relates them through one profile $g_N(y)$,
with $N\ge0$,
\begin{equation}
 \mathbf{d}_{N+1}^{\rm lin}
 =\frac{2g_N}{y^2}(\mathbf{a}^{(N)}\cdot\hat{\mathbf{r}})\hat{\mathbf{r}}
 +\frac{g_N'}y\left[\mathbf{a}^{(N)}-(\mathbf{a}^{(N)}\cdot\hat{\mathbf{r}})\hat{\mathbf{r}}\right] \ .
 \label{tube-linear-profile}
\end{equation}
Here primes denote $d/dy$. Equation~\eqref{tube-core-trace} gives
the formal linear force series
\begin{equation}
 \mathbf{F}_{\rm tubular}^{\rm linear}
 =\frac{q^2}{6\pi}\sum_{N=0}^{\infty}r_{\rm BI}^{N-1}g_N(0)\mathbf{a}^{(N)} \ .
 \label{tube-linear-trace}
\end{equation}
The order-one magnetic field vanishes under the same boundary conditions.
For higher orders, the radial part of the second curl equation in
\eqref{tube-full-recursion} gives
$\mathbf{b}_{N+2}^{\rm lin}=yg_N(y)(\mathbf{a}^{(N+1)}\times\hat{\mathbf{r}})/\sqrt{1+y^4}$.
Substitution in the electric equation gives
\begin{equation}
 \mathscr D_1^E g_0=\frac{y}{\sqrt{1+y^4}} \ ,\qquad
 \mathscr D_1^E g_1=0 \ ,\qquad
 \mathscr D_1^E g_{N+2}=\frac{y^2g_N}{\sqrt{1+y^4}}\quad(N\ge0) \ ,
 \label{tube-linear-main-recursion}
\end{equation}
where the operator $\mathscr D_1^E$ is such that  $\mathscr D_1^E\psi\equiv y[(y\psi)'/\sqrt{1+y^4}]'$.

\Needspace{5\baselineskip}
Each profile has two integration constants. The $1/y$ homogeneous mode produces a $y^{-3}$ displacement,
violating the imposed core bound and translating the source distribution.
Its coefficient therefore vanishes. The remaining constant is fixed by matching
the complete electric field to the retarded exterior field.
Appendix~\ref{app:tube-linear} gives the radial solutions and constructs
the retarded dipole fields from their exterior tails.

At $N=1$, a $y^0$ term in $\mathbf e_1$ would give an $r_{\rm BI}^{-1}\mathbf a$ contribution at fixed physical radius.
Matching to the finite Maxwell limit fixes $g_0(0)=-\eta$. At $N=2$, the retarded charge field's
$\tfrac23\dot{\mathbf{a}}$ term in the normalization of
\eqref{tube-field-series} fixes $g_1(0)=1$. At the next three orders, the retarded
data in \eqref{tube-linear-dipole-matching} give
$g_N(0)=\gamma_{N-1}$ for $N=2,3,4$, with
\begin{equation}
 \gamma_1=\frac{2\eta^3}{9}-\frac{\pi}{6\eta} \ ,\qquad
 \gamma_2=-\frac{2\eta^2}{3} \ ,\qquad
 \gamma_3=-\frac{8\eta^5}{135}
 +\frac{2\eta}{3}\left(\frac{31}{30}+\frac\pi6+\frac35\ln2\right) \ .
 \label{tube-gammas}
\end{equation}
The linear self-force through $r_{\rm BI}^3$ is therefore
\begin{equation}
 \left.\mathbf{F}_{\rm tubular}^{\rm linear}\right|_{N\le5}=-M_{\rm BI}\mathbf{a}+
 \frac{q^2}{6\pi}\left[\dot{\mathbf{a}}
 +r_{\rm BI}\gamma_1\ddot{\mathbf{a}}
 +r_{\rm BI}^2\gamma_2\mathbf{a}^{(3)}
 +r_{\rm BI}^3\gamma_3\mathbf{a}^{(4)}\right] \ .
 \label{tube-linear-result}
\end{equation}
The mass and the leading dissipative term agree with Feenberg's result~\cite{Feenberg},
but the coefficients differ starting at $r_{\rm BI}$.
Appendix~A gives the comparison in the same
units and normalization.

\subsection{Nonlinear Completion}
\label{sec:tube-nonlinear-calculation}

At each order we solve the angular sectors needed for the force and
higher-order sources, with the core and retarded conditions outlined
at the head of the previous subsection.
The nonlinear force terms through $r_{\rm BI}^3$ are such that
\bea\label{F_tubular_BI}
 \mathbf{F}_{\rm tubular}
 &=&\mathbf{F}_{\rm tubular}^{\rm linear}+\frac{q^2}{6\pi}\Big\{r_{\rm BI}\delta_1\mathbf{a}^2\mathbf{a}
 +r_{\rm BI}^2\big[A_2\mathbf{a}^2\dot{\mathbf{a}}
                  +B_2(\mathbf{a}\cdot\dot{\mathbf{a}})\mathbf{a}\big]\label{tube-nonlinear-force}\\
 &&+r_{\rm BI}^3\big[b_1\mathbf{a}^2\ddot{\mathbf{a}}
 +b_2(\mathbf{a}\cdot\ddot{\mathbf{a}})\mathbf{a}+b_3\dot{\mathbf{a}}^{\,2}\mathbf{a}
 +b_4(\mathbf{a}\cdot\dot{\mathbf{a}})\dot{\mathbf{a}}
                       +b_5(\mathbf{a}^2)^2\mathbf{a}\big]\Big\}+\cdots \ . \nonumber
\eea
Parity excludes axial vectors such as $\mathbf{a}\times\dot{\mathbf{a}}$
from the force.

\paragraph{Field order $N=2$.}
The field $\mathbf{d}_1$ can supply terms quadratic in $\mathbf{a}$ in (\ref{tube-full-recursion}). The matched $l=2$ displacement
in Appendix~\ref{app:tube-second-fields} has no dipole trace and hence
gives no additional force.

\paragraph{Field order $N=3$.}
The nonlinear sources contain $a_i\dot a_j$ and $a_i a_j a_k$.
Only the $l=1$ part proportional to $a_i a_j a_k$ contributes
to the nonlinear force, with vector structure $\mathbf{a}^2\mathbf{a}$.
The core and retarded conditions in Appendix~\ref{app:tube-third-fields}
fix its trace, giving
\begin{equation}
 \delta_1=-\frac{\eta^3}{9}+\frac{\pi}{12\eta}
 =-\frac{\gamma_1}{2} \ .
 \label{tube-delta1}
\end{equation}

\paragraph{Field order $N=4$.}
Matching the $l=1$ electric field in Appendix~\ref{app:tube-order4}
separately fixes the coefficients of $\mathbf{a}^2\dot{\mathbf{a}}$ and
$(\mathbf{a}\cdot\dot{\mathbf{a}})\mathbf{a}$, giving
\begin{equation}
 A_2=B_2=\frac23\eta^2=-\gamma_2 \ .
 \label{tube-order2}
\end{equation}

\paragraph{Field order $N=5$.}
Matching the five electric dipole projections in
Appendix~\ref{app:tube-order5} determines $b_1,\ldots,b_5$ separately.
The resulting coefficients satisfy
\begin{align}
 b_1&=-\gamma_3-\frac43b_5\ ,\qquad b_2=-\frac83b_5\ ,
 \label{tube-b1}\\
 b_3&=\frac32\gamma_3-\frac83b_5-2\eta\ ,\qquad
 b_4=-\gamma_3-\frac{16}{3}b_5 \ .
 \label{tube-other-b}
\end{align}
They are determined by a single new number,
\bea\label{tube-b5}
 b_5=-\frac{\eta^5}{18}
 +\frac{2\eta}{3}\left(\frac35\ln2+\frac\pi8+\frac{11}{240}
                                      -\frac{27}{20}G\right) \ ,
\eea
with Catalan's constant $G$,
\bea\label{Catalan}
G= \sum_{k\ge0}(-1)^k/(2k+1)^2\simeq 0.915965594177219\ldots \ .
\eea

\begin{table}[H]
\centering
\caption{Dimensionless coefficients of the tubular self-force, modulo the
universal factor $q^2/(6\pi)$. See (\ref{BI_mass}) and (\ref{Catalan}) for $\eta$ and $G$, respectively.}
\label{tab:tube-coefficients}
\vskip 3mm
\small
\renewcommand{\arraystretch}{1.35}
\setlength{\tabcolsep}{5pt}
\begin{tabular}{@{}cccr@{}}
\toprule
Order & Coefficients & Closed forms & Numerical Values\\
\midrule
$r_{\rm BI}$ & $\gamma_1$
 & $\dfrac{2\eta^3}{9}-\dfrac{\pi}{6\eta}$ & $1.133940\dots$\\[5pt]
 & $\delta_1$
 & $-\dfrac{\eta^3}{9}+\dfrac{\pi}{12\eta}$ & $-0.566970\dots$\\[5pt]
\addlinespace[5pt]
$r_{\rm BI}^2$ & $\gamma_2$ & $-\dfrac{2\eta^2}{3}$ & $-2.291728\dots$\\[5pt]
 & $A_2=B_2$ & $\dfrac{2\eta^2}{3}$ & $2.291728\dots$\\[5pt]
\addlinespace[5pt]
$r_{\rm BI}^3$ & $\gamma_3$
 & $-\dfrac{8\eta^5}{135}+\dfrac{2\eta}{3}
       \left(\dfrac{31}{30}+\dfrac\pi6+\dfrac35\ln2\right)$ & $1.140152\dots$\\[5pt]
 & $b_5$
 & $-\dfrac{\eta^5}{18}+\dfrac{2\eta}{3}
       \left(\dfrac35\ln2+\dfrac\pi8+\dfrac{11}{240}-\dfrac{27}{20}G\right)$
 & $-1.689539\dots$\\[5pt]
 & $b_1$ & $-\gamma_3-\dfrac43b_5$ & $1.112567\dots$\\[5pt]
 & $b_2$ & $-\dfrac83b_5$ & $4.505439\dots$\\[5pt]
 & $b_3$ & $\dfrac32\gamma_3-\dfrac83b_5-2\eta$ & $2.507519\dots$\\[5pt]
 & $b_4$ & $-\gamma_3-\dfrac{16}{3}b_5$ & $7.870726\dots$\\[5pt]
\bottomrule
\end{tabular}
\end{table}

\subsection{Covariant Form}

Consider any vector $V(\tau)$ defined over the worldline.
With the general decomposition $V^\mu=V^0u^\mu+\mathbf V^ie_i^\mu$
and the triad property $\dot e_i^\mu=a_i u^\mu$, we find that
\bea \label{tube-cov-dictionary}
 \frac{dV^\mu}{d\tau} \quad\longleftrightarrow\quad
 (\dot V^0+\mathbf{a}\cdot\mathbf{V} ,\ \dot{\mathbf{V}}+V^0\mathbf{a}) \ ,
\eea
while the orthonormal triad defines the projector $\Delta$,
\bea
 \Delta^\mu{}_{\nu}  =\sum_i e_i^\mu e_{i\nu}=\delta^\mu{}_{\nu}+u^\mu u_\nu
 \ ,
\eea
which separates out the part transverse to $u$ since $u^2=-1$.

Repeated application of \eqref{tube-cov-dictionary} gives the covariant
form of the self-force as
\begin{align}
 F_{\rm tubular}^\mu={}&-M_{\rm BI}a^\mu+
 \frac{q^2}{6\pi}\Big\{\Delta^\mu{}_{\nu}\dot a^\nu
 +r_{\rm BI}\gamma_1\left(\Delta^\mu{}_{\nu}\ddot a^\nu
                 -\frac32a^2a^\mu\right)\nonumber\\
 &+r_{\rm BI}^2\Big[\gamma_2\Delta^\mu{}_{\nu}a^{(3)\nu}
 +(A_2-\gamma_2)a^2\Delta^\mu{}_{\nu}\dot a^\nu
 +(B_2-5\gamma_2)(a\cdot\dot a)a^\mu\Big]\nonumber\\
 &+r_{\rm BI}^3\Big[\gamma_3\Delta^\mu{}_{\nu}a^{(4)\nu}
 +(b_1-\gamma_3)a^2\Delta^\mu{}_{\nu}\ddot a^\nu
 +(b_2-9\gamma_3)(a\cdot\ddot a)a^\mu\nonumber\\
 &\hspace{17mm}+(b_3-8\gamma_3)(\dot a\cdot\dot a)a^\mu
 +(b_4-7\gamma_3)(a\cdot\dot a)\Delta^\mu{}_{\nu}\dot a^\nu\nonumber\\
 &\hspace{17mm}+(b_5-b_1-b_2+b_3+\gamma_3)(a^2)^2a^\mu\Big]\Big\}+\cdots \ .
 \label{tube-covariant-result}
\end{align}
Note that $u_\mu\Delta^\mu{}_{\nu}=0$ and $u\cdot a=0$ imply
$u_\mu F_{\rm tubular}^\mu=0$ at each displayed order.
It is important to remember that we did not add new terms for the sake of
the relativistic completion but merely reorganized terms to fit the general
covariance.

Before closing the section, let us note that under uniform
acceleration, that is, for $\dot a^\mu=a^2u^\mu$,
the ALD force vanishes as is familiar from Maxwell theory, leaving behind
\begin{equation}
 F_{\rm tubular}^\mu\biggr|_{\rm uniform}
 =\left[-M_{\rm BI}-\frac{q^2}{6\pi}\frac{r_{\rm BI}\gamma_1}{2}a^2
       +\frac{q^2}{6\pi}r_{\rm BI}^3b_5(a^2)^2\right]a^\mu+\cdots \ .
 \label{tube-uniform-check}
\end{equation}
This is one of the qualitative deviations from the Maxwell theory,
where the self-force vanishes identically under such an idealized motion.
However, this fact is still consistent with the hyperbolic motions as
static trajectories in the Rindler coordinate system. All that happens is that, of the energy inflow due to the force driving the acceleration,
a nonzero fixed fraction determined by $r_{\rm BI}a$ flows into the electromagnetic
field energy instead of being used for the particle acceleration.

\section{Bremsstrahlung}
\label{sec:bi-bremsstrahlung}

The previous section extracted the self-force from the stress flux through a
shrinking tube around the point charge. When we did this, it was critical that
we use solutions that obey the right boundary condition not only at the origin
but also in the asymptotic region, to make sure that the usual retarded
nature of classical field propagation is correctly implemented. This means that
we are already equipped to follow the same matched field into the far zone to 
calculate the energy and momentum carried away by its  radiative field. We 
record the results here for completeness and also for the self-consistency check
of the self-force derived. 

We will see that the radiative energy-momentum takes a far simpler form,
relative to the work done by self-force. This is because part of the latter
is reversible, which manifests as a time derivative piece in $F_{\rm self}$, 
to be denoted $dp_{\rm b}/d\tau$. The so-called ``bound momentum'' $p_{\rm b}$ here is an indispensable
part of the comparison between the self-force and the Bremsstrahlung. Once we isolate 
the energy-momentum radiated, $\mathcal R$ in (\ref{brems-flux-definition}) below, 
the remainder of $F_{\rm self}= F_{\rm tubular}$ must  be organized into  
a total $\tau$-derivative, which by itself forces relations among coefficients as defined 
by (\ref{F_tubular_BI}). We will close the section by observing how the numbers 
we computed in Section~\ref{sec:tubular} match these reversibility criteria
precisely, which completes the comparison to the given order.

At large observation radius $R$, the field is weak, so the radiated
energy and momentum are determined by the Maxwell stress tensor of its
$1/R$ radiation field. We evaluate the radiation field from
\eqref{tube-retarded-length-expansion} on the null cuts
$t-R=z^0(\tau)-\hat{\mathbf r}\cdot\mathbf z(\tau)$ of the physical
worldline, with observation direction $\hat{\mathbf r}$.
Let $\mathbf E_\perp$ denote the $1/R$ part of
$\mathbf E-\hat{\mathbf r}(\hat{\mathbf r}\cdot\mathbf E)$.
At an instantaneous rest event, $d(t-R)/d\tau=1$, so the radiated energy
and momentum per unit source proper time are
\bea
 (P_{\rm rad},\boldsymbol{\mathcal R})
 &=&\lim_{R\to\infty}\frac{R^2}{\bbe^2}
       \int(1,\hat{\mathbf r})|\mathbf E_\perp|^2\,d\Omega\ ,\cr\cr
 \mathcal R^\mu&=&P_{\rm rad}u^\mu+\mathcal R^i e_i^\mu \ .
 \label{brems-flux-definition}
\eea
As with any 4-momentum, the time-component represents the energy, so
$P_{\rm rad}$ represents energy outflow per unit time, or the power.

For Maxwell, we recall
that
\bea
 P_{\rm rad}\biggr\vert_{\rm Maxwell} =\frac{q^2}{6\pi}\mathbf a^2 \ .
\eea
The BI correction to the exterior radiation field starts at $r_{\rm BI}^2$.
Its contribution to the flux through $r_{\rm BI}^3$ is therefore the
interference with the charge radiation. Angular integration gives
\begin{align}
 P_{\rm rad}=\frac{q^2}{6\pi}\Big\{&\mathbf a^2
 -\frac{2\eta^2r_{\rm BI}^2}{3}
  \left[\mathbf a\cdot\ddot{\mathbf a}-\frac25(\mathbf a^2)^2\right]
 \nonumber\\
 &+\frac{2\eta r_{\rm BI}^3}{3}
  \left[2\mathbf a\cdot\mathbf a^{(3)}
       -\frac45\mathbf a^2(\mathbf a\cdot\dot{\mathbf a})\right]
 \Big\}+\cdots \ ,
 \label{brems-power}
\end{align}
and
\begin{align}
 \boldsymbol{\mathcal R}=\frac{q^2}{6\pi}\Big\{&
 -\frac{2\eta^2r_{\rm BI}^2}{15}
  \left[(\mathbf a\cdot\dot{\mathbf a})\mathbf a
        +3\mathbf a^2\dot{\mathbf a}\right]
 \nonumber\\
 &+\frac{2\eta r_{\rm BI}^3}{15}
  \left[7\mathbf a^2\ddot{\mathbf a}
       -(\dot{\mathbf a}^2+\mathbf a\cdot\ddot{\mathbf a})\mathbf a
       +3(\mathbf a\cdot\dot{\mathbf a})\dot{\mathbf a}\right]
 \Big\}+\cdots \ .
 \label{brems-momentum}
\end{align}
For uniform acceleration, $\dot{\mathbf a}=0$ gives
\begin{equation}
 P_{\rm rad}=\frac{q^2}{6\pi}
 \left[\mathbf a^2+\frac{4\eta^2r_{\rm BI}^2}{15}(\mathbf a^2)^2\right]
 +\cdots \ ,\qquad \boldsymbol{\mathcal R}=0+\cdots \ .
 \label{brems-uniform}
\end{equation}

In a fixed inertial frame, the net emitted four-momentum is the proper-time
integral $\Delta p_{\rm rad}^\mu=\int\mathcal R^\mu\,d\tau$, whose time
component is the emitted energy.
For inertial endpoints with
$\mathbf a=\dot{\mathbf a}=\ddot{\mathbf a}=0$, integration by parts in
\eqref{brems-power}--\eqref{brems-momentum} gives
\begin{equation}
 \begin{aligned}
 \Delta p_{\rm rad}^\mu=\frac{q^2}{6\pi}\int d\tau\,\Big\{&
 \mathbf a^2u^\mu\\
 &+\frac{2\eta^2r_{\rm BI}^2}{3}
 \big[[\dot{\mathbf a}^2+(\mathbf a^2)^2]u^\mu
             +2(\mathbf a\cdot\dot{\mathbf a})a^\mu\big]\\
 &+2\eta r_{\rm BI}^3\dot{\mathbf a}^2a^\mu\Big\}+\cdots \ .
 \end{aligned}
 \label{brems-integrated-expansion}
\end{equation}
For slow motion about rest, the energy through $r_{\rm BI}^3$ at
quadratic order in the motion amplitude is
\begin{equation}
 \Delta p_{\rm rad}^0\simeq\frac{q^2}{6\pi}
 \left[\int\mathbf a^2\,d\tau
 +\frac{2\eta^2r_{\rm BI}^2}{3}\int\dot{\mathbf a}^{\,2}\,d\tau\right] \ .
 \label{brems-quadratic-energy}
\end{equation}

As is already well known with the ALD force, the radiated energy and the work
done by the self-force need not match each other in real time. This
is famously because there is one more component to the energy budget, namely
the energy of the Coulombic field moving with the charged particle. The
work done on the particle by the self-force transfers energy to this near field first, some of which is eventually radiated away to the asymptotic region. The
famous historical puzzle of a uniformly accelerated particle, for which
radiative energy and the self-force mismatch most starkly, resolves itself
once we take into account this near-field energy reservoir.\footnote{In the case of Maxwell theory, the latter is divergent and can easily muddle the naive
energy budget; in Born--Infeld theory with its finite self-energy, however,
one should be able to discuss the entire energy balance without such extraneous issues.}

In the particle equation of motion, this near-field energy reservoir manifests
as a total time derivative, as in
\begin{equation}
 F_{\rm tubular}^\mu+M_{\rm BI}a^\mu
 +\frac{dp_{\rm b}^\mu}{d\tau}+\mathcal R^\mu=0 \ .
 \label{brems-balance}
\end{equation}
The work done by $dp_{\rm b}/d\tau$ is reversible; for a motion
that starts and ends inertially, the integrated power due to it
vanishes identically.

On the instantaneous-rest hyperplane $\Sigma_\tau$, the bound momentum
$p_{\rm b}^\mu$ may be computed as
\begin{equation}
 p_{\rm b}^\mu
 =\int_{\Sigma_\tau}T^{0\mu}\,d^3x-M_{\rm BI}u^\mu
       -\int_{-\infty}^{\tau}\mathcal R^\mu(\tau')\,d\tau' \ .
 \label{brems-bound-definition}
\end{equation}
Integrating the matched fields gives the bound-momentum components
$p_{\rm b}^0=-u_\mu p_{\rm b}^\mu$ and
$p_{\rm b}^i=e_{i\mu}p_{\rm b}^\mu$ as
\begin{align}
 p_{\rm b}^0=\frac{q^2}{6\pi}\Big\{&
 \frac12r_{\rm BI}\gamma_1\mathbf a^2
 +r_{\rm BI}^3\big[(\gamma_3-\tfrac43\eta)\mathbf a\cdot\ddot{\mathbf a}
 +(\tfrac23\eta-\tfrac12\gamma_3)\dot{\mathbf a}^2
 -b_5(\mathbf a^2)^2\big]\Big\}+\cdots \ ,\label{brems-bound-energy}\\
 \mathbf p_{\rm b}=\frac{q^2}{6\pi}\Big\{&
 -\mathbf a-r_{\rm BI}\gamma_1\dot{\mathbf a}
 +r_{\rm BI}^2\gamma_2
       [-\ddot{\mathbf a}+\tfrac25\mathbf a^2\mathbf a]\nonumber\\
 &+r_{\rm BI}^3\big[-\gamma_3\mathbf a^{(3)}
  +(\gamma_3+\tfrac43b_5-\tfrac{14}{15}\eta)\mathbf a^2\dot{\mathbf a}
 \nonumber\\
 &\hspace{19mm}+(-\gamma_3+\tfrac83b_5+\tfrac{22}{15}\eta)
          (\mathbf a\cdot\dot{\mathbf a})\mathbf a\big]\Big\}+\cdots \ ,
 \label{brems-bound-momentum}
\end{align}
with $\gamma_1,\gamma_2,\gamma_3,b_5$ as given in Section~\ref{sec:tubular}.

As mentioned at the top of the section, the fact that the difference between
the self work and the radiative energy must be reversible demands relations among
the coefficients in (\ref{F_tubular_BI}). For instance, at order $r_{\rm BI}^3$, 
this energy balance condition requires the relations:
\begin{equation}
 \begin{aligned}
 \gamma_3+2b_1+b_2-b_4&=0 \ ,\\
 6b_1+b_2+2b_3-3b_4&=-4\eta \ ,\\
 b_1-b_4-4b_5&=0 \ .
 \end{aligned}
 \label{brems-balance-check}
\end{equation}
Substituting the self-force coefficients obtained in
\eqref{tube-b1}--\eqref{tube-other-b} verifies all three relations.

One should be mindful that the Bremsstrahlung computed above
is not always nonnegative order by order, although the general
expression (\ref{brems-flux-definition}) is manifestly
nonnegative. This is an unavoidable artifact of a power expansion.
For instance, we find that the $r_{\rm BI}^3$-order expression is not of definite
sign, yet by having an appropriate $r_{\rm BI}^6$ piece, this can be
easily remedied.

\section{Lorentz vs Tubular: What Does $S^{\rm self}_{\rm in}$ Do?}
\label{sec:lorentz-tubular}

We have seen how the Bremsstrahlung and the self-force results above 
go hand in hand, telling us that the tubular self-force we have computed
starting from (\ref{shift}) is the right answer to the original question.
There is something odd about (\ref{shift}), however, in that it arose from varying only the $S^{\rm self}_{\rm out}$ part of the on-shell action  $S^{\rm self}$. 

Moreover, if we consider it the most general
form of the self-force, we might say that the surprise is how the PV-regulated Maxwellian
self-force as computed by Coleman came entirely from the Lorentz force. At the same time,
Teitelboim~\cite{Teitelboim1970,Teitelboim1971a,Teitelboim1971b} had addressed the self-force problem of the Maxwell theory in yet another
manner, where he took the singular point-like source and the divergent energy-momentum
at face value, and offered the Lorentz self-force and the tubular self-force as  
alternatives to each other.\footnote{One caveat for Teitelboim's
treatment is that
the ``Lorentz force'' he computed was not an honest one, having
been evaluated over a fictitious two-sphere surrounding the point-like charge as if the charge
were distributed over the sphere;  he did not smear
the charge by replacing the delta-function source by a ball or by a sphere.
One should  regard  the ``Lorentz force'' offered by Teitelboim in his third paper~\cite{Teitelboim1971b}
merely as a mathematical rewriting of the tubular momentum transfer
in Refs.~\cite{Teitelboim1970,Teitelboim1971a}. }

One thing Teitelboim reminds us of is a general identity that would have motivated
such a rewriting.
For any reasonable electrodynamics, linear or nonlinear, one finds
\bea\label{EM_balance}
F_{\rm tubular}\biggr\vert_{\mathbb S^2}=F_{\rm Lorentz}\biggr\vert_{\mathbb B^3} + \frac{dp_{\rm EM}}{d\tau}\biggr\vert_{\mathbb B^3} \ ,
\eea
where $ p_{\rm EM}$ is the momentum density of the electromagnetic field integrated 
over $\mathbb B^3$. The identity (\ref{EM_balance}), which arises from the combination of the
energy-momentum conservation and the electromagnetic field equation, holds
in general as long as all physical quantities are smooth.

For any fully relativistic electrodynamics coupled to an external current, the conservation law reduces to
\bea\label{EM_balance_local}
-\partial_\mu T^\mu{}_{\nu} = -\left[(\partial_\mu \cP^{\mu\alpha})\cF_{\nu\alpha}
 +\cP^{\mu\alpha}\partial_\mu \cF_{\nu\alpha}
 +\partial_\nu\mathcal L\right] =\cF_{\nu\alpha}j^\alpha \ ,
\eea
where we used
\bea
 \cP^{\mu\nu}\equiv-2\frac{\partial\mathcal L}{\partial \cF_{\mu\nu}}\ , \qquad \partial_\nu \cP^{\mu\nu}=j^\mu \ ,
\eea
on par with the BI example in Section~\ref{sec:bi-maxwell} and the Bianchi identity for $\cF$.
Integrating (\ref{EM_balance_local}) over $\mathbb B^3$ leads us to (\ref{EM_balance}).
The failure of $T$ to be conserved on its own is simply because this $T$  does not include the contribution of the charged matter field and this manifests via the Lorentz
force on the right. As such, this equality as a whole is the conservation law.

Nevertheless, for the two examples in this note, Coleman's Maxwell and our
Born--Infeld with their point-like charges, the identity (\ref{EM_balance}) appears to fail. 
For the case of Coleman, only $F_{\rm Lorentz}$ survives. Due to
finiteness of $(\cF_{\rm self})^2$, neither $F_{\rm tubular}$ nor the momentum inside
$p_{\rm EM}$ survives the infinitesimal tubular limit, yet we have computed finite
$F_{\rm Lorentz}$ at the point-charge source, leading to a violation of (\ref{EM_balance}).
In this case the issue is simpler: the PV-regulated field equation does not
follow from the action used to compute the energy-momentum tensor, so a violation
of the conservation law (\ref{EM_balance_local}) is hardly a surprise. The energy-momentum
conservation does not hold for Coleman's regulated solution, in other words, so that
the question shifts to how we should view this regulator and how we must
use the regulated solution.

Things are  less obvious when it comes to the BI analysis with a point-like charge,
for which the naive Lorentz force vanishes, despite how the energy-momentum conservation 
has to hold strictly. For this, we should recall that the point charge with its
delta function support should be considered in the sense of ``distribution.''
Even with Coleman's treatment of Maxwell, this was needed to justify the
angular average that deals with the potential ambiguity at the origin. By
approaching the problem more carefully this way, we shall finally arrive at what the
variation of $S^{\rm self}_{\rm in}$ does, or more precisely how it does not
add a new term for self-force beyond $F_{\rm tubular}$. 

To say that the point charge should be considered a distribution is equivalent
to replacing it by a ball of smeared charge and taking the infinitesimal limit
in the end. In this case, everything is regular and smooth so that (\ref{EM_balance_local})
should hold and by inference (\ref{EM_balance}) also should hold for
$\mathbb S^2_{\rm R}$ surrounding the ball of radius $R$. Indeed one can track the computation for
BI theory for this more involved case, well enough to see that, if one again takes
the infinitesimal ball limit, $R\rightarrow 0$,
\bea
\lim_{R\rightarrow 0} F_{\rm tubular}\biggr\vert_{\mathbb S^2_R}=\lim_{R\rightarrow 0} F_{\rm Lorentz}\biggr\vert_{\mathbb B^3_R} \ ,
\eea
with $dp_{\rm EM}/d\tau$ dropping out upon the same infinitesimal limit. The
field near the smeared charge source does not converge uniformly to that of
a point charge; the two disagree inside the smeared ball charge, but this
difference becomes undetectable and thus drops out if we start naively with a
point source from the beginning. 

In the context of variation of $S^{\rm self}_{\rm in}$ in the limit of 
$R=\epsilon\rightarrow 0$, this last computation tells us 
\bea
\delta S^{\rm self}_{\rm in} \quad\Rightarrow \quad 
\lim_{\epsilon \rightarrow 0} 
\left(F_{\rm Lorentz}\biggr\vert_{\mathbb B^3_\epsilon}
-F_{\rm tubular}\biggr\vert_{\mathbb S^2_\epsilon}\right) = 0\ ,
\eea
for which it is central that the point charge must be treated properly
as a distribution. Recall that we initially chose the sign of $F_{\rm tubular}$ 
at $\mathbb S^2_\epsilon$ to fit the boundary term from the variation of $S^{\rm self}_{\rm out}$; 
the corresponding boundary term from $S^{\rm self}_{\rm in}$ is obviously the same except for the sign flip. 
This allows us to conclude that (\ref{shift}) is indeed the entire self-force 
that results from varying 
\bea
S^{\rm self}=S^{\rm self}_{\rm out}+S^{\rm self}_{\rm in}=S^{\rm self}_{\rm BI}+S^{\rm self}_{\rm minimal}\ ,
\eea
provided that we treat the point charge properly as a distribution. 

If we had approached the problem as a momentum-transfer question as in (\ref{EM_balance})
from the beginning, it would have remained unclear  what $\mathbb S^2$ should be used for 
estimating the tubular  force on ``the particle'' given the mass renormalization
of the charged particle due to the near field. For a point-like charge in BI theory, 
on the other hand, the action principle safeguards us against such a conceptual 
``red herring,'' leading us to (\ref{shift}) unambiguously. 

This also allowed us to avoid   a computational and technical headache 
associated with the point charge. The earlier observation that the Lorentz 
self-force vanishes was not quite correct, as it ignored the true nature of
the point charge as a distribution and the subsequent non-uniform behavior 
of $\cF$ at the origin. By invoking the tubular self-force from the variation 
of $S^{\rm self}_{\rm out}$, we have essentially sidestepped the subtlety 
at the charge location. 

\section{Summary}

One could have started this investigation in part motivated by the naive thought that,
perhaps, Bremsstrahlung of nonlinear electrodynamics may reduce to that of Maxwell theory.
After all, BI electrodynamics reduces reliably to Maxwellian electrodynamics in the
weak field limit and even more importantly the usual wave solution of Maxwell theory
remains a solution of the BI theory. If we had taken these thoughts at face value, 
we would have anticipated no $\ell$ corrections in the BI Bremsstrahlung relative
to that of Maxwell and, perhaps, the same with the self-force as well. 

This naivete proves itself wrong quickly even without a detailed estimate.
Although the asymptotic wave propagation does reduce to that of Maxwellian linear
dynamics, what drives these waves are dipoles and higher moments that must be
computed near the charge core where the nonlinearity is strongest. Section~\ref{sec:bi-bremsstrahlung} on the BI Bremsstrahlung demonstrates this amply and directly, which
was also foreshadowed by the self-force evaluation. In this note we have
computed the two effects, near the charge core and in the asymptotic region,
respectively, and found a perfect agreement between the two  to the
$\ell^3$ order computed.

One quick generalization can be found in the so-called Born--Infeld ModMax (BIMM)
theory~\cite{ModMax}, which is a one-parameter 
generalization of BI with the electromagnetic self-duality intact. 
Here again, absence of extra regularization would mean that, 
regardless of details  of the model, the electric displacement $\mathbf D$ retains
the same old $\sim 1/r^2$ peak at the point-like charge source
while its $\mathbf E$ counterpart remains bounded from above everywhere.
Since this $\mathbf D^i=\cP^{0i}$ arises from variation
of the action with respect to $\mathbf E^i=\cF^{0i}$, a tubular force of the form
(\ref{shift1}) would be again responsible for the self-force. Appendix B
records how the final form of the self-force to the $\ell^3$ order is essentially
the same as that of the BI results to the given order, although this coincidence
is expected to fail at higher orders. 

The usual ALD force with no dimensionful parameter attached is fairly universal and can be
apparently derived with less attention given to the procedure.
In a sense, we are again seeing the same universality in BI and BIMM theories. The difference is 
that $\ell$ is no longer a regularization device in these theories so that we end up 
with a subleading derivative expansion weighted by positive powers of $\ell$.
BI theory as well as BIMM theories are one set of examples where every single term in the
derivative expansion of the self-force, including the mass renormalization term, is
finite and carries well-defined physical meaning of its own.

We emphasize again how the tubular contribution from the on-shell action 
is generically how the self-force should enter the story in finite and 
nonlinear electrodynamics.
A sufficient condition for this is that $\cF$ is controlled by nonlinearity
to be finite at the point-charge location. With the analog of $\cF_{\rm com}$
finite everywhere, the subsequent perturbation due to acceleration is likely
to be subleading in small $r$ and to vanish at the charge location if the
charge is treated as truly point-like. In a way, these nonlinear
and finite electrodynamics allow the cleanest derivation of self-force,
thanks to how $S^{\rm self}$---on-shell for
the field variables---is amenable to a variational problem with respect to the particle
position. This fact that the on-shell action 
$S^{\rm self}$ offers an unambiguous and clean derivation of the self-force 
is one general lesson worthwhile to remember from the exercise we offer.

\section*{Acknowledgement}

We relied on GPT-5.6 Sol and Claude Opus 5.5 for many detailed corners of this work
such as solving the nonlinear field equations order by order, and performing
the tubular integration. We gathered these detailed middle steps into
Appendix C, concentrating on physics in the main text. The comparison
to Feenberg's old results in Appendix A was also entrusted to ChatGPT. PY thanks 
his graduate advisor, John P. Preskill, who, decades ago, handed over a copy of 
Coleman's treatise, and also Doojin Kim who, as an undergraduate and again decades ago,
unearthed Feenberg's paper. This work is supported by KIAS Individual Grants (KL, PG096201; PY, PG005705).

\appendix

\section{Comparisons to Feenberg}

\paragraph{Notation and unit conversion dictionary.} As in the main text, the only building blocks are the charge $q=\bbe n$, where $\bbe$ and $n$ are the
coupling and the integer charge of the action and $q^2=\bbe^2n^2$ is used as it is, together with
\begin{equation}
r_{\rm BI}^2=\frac{\bbe^2|n|}{4\pi}\,\ell^2\ ,\qquad
M_{\rm BI}=\eta\,\frac{q^2}{6\pi r_{\rm BI}} \ ,\qquad
\eta=\int_0^\infty\frac{dy}{\sqrt{1+y^4}}=\frac{\Gamma(1/4)^2}{4\sqrt\pi}
 \ .
\label{feenberg-notation}
\end{equation}
Here $\ell$ is the BI length of the action ($b=\ell^{-2}$), $y=r/r_{\rm BI}$, and $M_{\rm BI}$ is the electrostatic field energy of the
core. We set $c=1$.

Feenberg's fields are in Gaussian units, whereas we use Heaviside--Lorentz units,
with a factor of $4\pi$ difference in the definition of the electric charge squared.
We denote Feenberg's quantities by a sub(super)script ``F''. The complete conversion
dictionary is
\begin{equation}
q_{\rm F}=\frac{q}{\sqrt{4\pi}} \ ,\qquad \mathbf D_{\rm F}=\bbe\sqrt{4\pi}\,\mathbf D\ ,\qquad
\mathbf E_{\rm F}=\frac{\sqrt{4\pi}}{\bbe}\,\mathbf E \ ,\qquad b_{\rm F}=\frac{\sqrt{4\pi}}{\bbe\ell^2} \ ,
\label{feenberg-dict}
\end{equation}
which also gives
\begin{equation}
\frac{2q_{\rm F}^2}{3}=\frac{q^2}{6\pi} \ .
\label{feenberg-q}
\end{equation}
Feenberg's core radius $r_0$ is the same as our $r_{\rm BI}$ of \eqref{feenberg-notation}. Everything below has been transcribed into
these conventions.

\paragraph{Feenberg's construction.} Feenberg's static fields, in Gaussian units, are
\begin{equation}
\mathbf D_{\rm F0}(\mathbf r)=\frac{q_{\rm F}}{r^2}\,\hat{\mathbf r}\ ,\qquad
\mathbf E_{\rm F0}(\mathbf r)=\frac{q_{\rm F}}{\sqrt{r^4+r_{\rm BI}^4}}\,\hat{\mathbf r} \ .
\label{feenberg-static}
\end{equation}
He defines a distributed charge density from the divergence of $\mathbf E_{\rm F0}$, not from that of $\mathbf D_{\rm F0}$:
\begin{equation}
\varrho_{\rm F}(r):=\frac{1}{4\pi}\,\nabla\!\cdot\mathbf E_{\rm F0}=\frac{q_{\rm F}}{2\pi r_{\rm BI}^3\,y(1+y^4)^{3/2}}\ ,\qquad y=\frac{r}{r_{\rm BI}} \ .
\label{feenberg-rho}
\end{equation}
He then uses it to construct
\begin{equation}
\mathbf F=\int\varrho_{\rm F}(\mathbf r)\,\mathbf D_{\rm F}(\mathbf r)\,d^3\mathbf r \ ,
\label{feenberg-F}
\end{equation}
and takes this as the self-force in the instantaneous rest frame. However familiar this integral may look, it is not the
Lorentz force: the roles of $\mathbf D$ and $\mathbf E$ are reversed. Nor does it give the tubular self-force of the main text. This is where
Feenberg's treatment went astray.

Because $\varrho_{\rm F}$ is spherically symmetric, only the angular average of the retarded displacement enters
\eqref{feenberg-F}:
\begin{equation}
\big\langle\mathbf D_{\rm F,ret}(\mathbf r,t)\big\rangle_\Omega=q_{\rm F}\Big[-\frac{2}{3r}\,\mathbf a+\frac23\,\dot{\mathbf a}-\frac{r}{3}\,\ddot{\mathbf a}
+\frac{r^2}{9}\,{\mathbf a}^{(3)}-\frac{r^3}{36}\,{\mathbf a}^{(4)}+\cdots\Big] \ .
\label{feenberg-Dret}
\end{equation}
The key step is the removal of $\mathbf D_{\rm F0}$, which is both singular and ambiguous at the origin. The integration
then gives Feenberg's self-force,
\begin{align}
\mathbf F^{\rm F}_{\rm self}={}&-\frac{4q_{\rm F}^2}{3r_{\rm BI}}\,\cG(0,3)\,{\mathbf a}+\frac{4q_{\rm F}^2}{3}\,\cG(1,3)\,\dot{\mathbf a}
-\frac{2q_{\rm F}^2r_{\rm BI}}{3}\,\cG(2,3)\,\ddot{\mathbf a}\cr
&+\frac{2q_{\rm F}^2r_{\rm BI}^2}{9}\,\cG(3,3)\,{\mathbf a}^{(3)}-\frac{q_{\rm F}^2r_{\rm BI}^3}{18}\,\cG(4,3)\,{\mathbf a}^{(4)}+\cdots \ ,
\label{feenberg-result}
\end{align}
with the coefficients given by the integrals
\begin{equation}
\cG(m,n)\equiv\int_0^\infty\frac{y^m\,dy}{(1+y^4)^{n/2}} \ .
\label{feenberg-G}
\end{equation}

\paragraph{Comparison of self-force results.}

The first two terms are sensible. With $\cG(0,3)=\Gamma(1/4)^2/(8\sqrt\pi)=\eta/2$, Feenberg's inertial coefficient is exactly the
electrostatic core energy of the main text,
\begin{equation}
\frac{4q_{\rm F}^2}{3r_{\rm BI}}\,\cG(0,3)=\eta\,\frac{q^2}{6\pi r_{\rm BI}}=M_{\rm BI} \ .
\label{feenberg-mass}
\end{equation}
With $\cG(1,3)=\frac12$, the next, dissipative term $\propto\dot{\mathbf a}$ also agrees with ours
and coincides with that of Maxwell theory. For the comparison of the rest, we need five
integrals. Given $\cG(m,3)=\frac14B\big(\frac{m+1}4,\frac32-\frac{m+1}4\big)$, we find
\begin{equation}
\cG(0,3)=\cG(4,3)=\frac{\eta}{2} \ ,\qquad \cG(1,3)=\cG(3,3)=\frac12 \ ,\qquad \cG(2,3)=\frac{\pi}{4\eta} \ .
\label{feenberg-int}
\end{equation}
Here $\cG(0,3)=\cG(4,3)$ follows from integrating $d[y(1+y^4)^{-1/2}]/dy=(1-y^4)(1+y^4)^{-3/2}$, and their sum is $\eta$.
$\cG(1,3)$ and $\cG(3,3)$ are elementary, while $\cG(2,3)=\Gamma(3/4)^2/(2\sqrt\pi)=\pi/(4\eta)$ through Euler's reflection
formula.

 Using \eqref{feenberg-q}, Eq.~\eqref{feenberg-result} becomes\footnote{
With $\eta=1.8540746773\ldots$ from \eqref{feenberg-notation}, this reproduces the numerical result quoted in the paper that
computed linear terms only~\cite{Feenberg},
\begin{equation*}
\mathbf F^{\rm F}_{\rm self}=-M_{\rm BI}\,{\mathbf a}+\frac{q^2}{6\pi}\Big[\dot{\mathbf a}-0.424\,r_{\rm BI}\,\ddot{\mathbf a}+0.167\,r_{\rm BI}^2\,{\mathbf a}^{(3)}
-0.077\,r_{\rm BI}^3\,{\mathbf a}^{(4)}+\cdots\Big] \ .
\end{equation*}
Direct evaluation gives $\pi/(4\eta)=0.4236\ldots$,
which rounds to $0.424$ rather than the $0.422$ printed in the paper, most likely a simple typo.}
\begin{equation}
\mathbf F^{\rm F}_{\rm self}=-M_{\rm BI}\,{\mathbf a}+\frac{q^2}{6\pi}\Big[\dot{\mathbf a}-\frac{\pi}{4\eta}\,r_{\rm BI}\,\ddot{\mathbf a}+\frac16\,r_{\rm BI}^2\,{\mathbf a}^{(3)}
-\frac{\eta}{24}\,r_{\rm BI}^3\,{\mathbf a}^{(4)}+\cdots\Big] \ .
\label{feenberg-eta}
\end{equation}
We compare this with the leading nonrelativistic part of our fully covariant self-force
\begin{equation}
\mathbf F^{\rm linear}_{\rm tubular}=-M_{\rm BI}\,{\mathbf a}+\frac{q^2}{6\pi}\Big[\dot{\mathbf a}+\gamma_1\,r_{\rm BI}\,\ddot{\mathbf a}+\gamma_2\,r_{\rm BI}^2\,{\mathbf a}^{(3)}
+\gamma_3\,r_{\rm BI}^3\,{\mathbf a}^{(4)}+\cdots\Big] \ ,
\label{tube-linear}
\end{equation}
whose coefficients inside the square bracket are listed in the last column of Table~\ref{tab:feenberg}.
After the leading dissipative term, the higher
motion-linear coefficients disagree. The culprit is \eqref{feenberg-F}, which may
have been an ad hoc device meant to get around the vanishing Lorentz ``self-force'' that we have noted in the main text.

\begin{table}[ht]

\caption{The leading coefficients for the self-force in \eqref{feenberg-eta} and
\eqref{tube-linear}, with $c=1$, modulo the universal factor $q^2/(6\pi)=2q_{\rm F}^2/3$. }
\label{tab:feenberg}
\vskip 3mm\renewcommand{\arraystretch}{2.0}
\small\begin{center}
\begin{tabular}{ccc}
leading linear term  & Feenberg & tubular self-force\\\midrule
$\dot{\mathbf a}$ & $1$ & $1$\\
$r_{\rm BI}\,\ddot{\mathbf a}$ & $-\dfrac{\pi}{4\eta}$ & $\dfrac29\eta^3-\dfrac{\pi}{6\eta}$\\
$r_{\rm BI}^2\,{\mathbf a}^{(3)}$ & $\dfrac16$ & $-\dfrac23\eta^2$\\
$r_{\rm BI}^3\,{\mathbf a}^{(4)}$ & $-\dfrac{\eta}{24}$ & $-\dfrac{8}{135}\eta^5+\dfrac23\eta\Big(\dfrac{31}{30}+\dfrac\pi6+\dfrac35\ln2\Big)$ \\
\end{tabular}
\end{center}
\end{table}

\section{Self-Force with Born--Infeld ModMax}
\label{app:bimm}

The BIMM theory~\cite{ModMax} generalizes BI by adding one dimensionless parameter $\gamma$.
The Lagrangian density is
\begin{equation}
 \mathcal L_\gamma=\frac{1}{\bbe^2\ell^4}\left[
 1-\sqrt{1+\frac{\ell^4}{2}
 \left(I\cosh\gamma-\sqrt{I^2+J^2}\sinh\gamma\right)
 -\frac{\ell^8J^2}{16}}\right] \ .
 \label{bimm-lagrangian}
\end{equation}
It recovers BI at $\gamma=0$ while its $\ell\to0$ limit is called the ModMax theory~\cite{ModMax,Kosyakov}.
We shall assume $\gamma\ge 0$ in order to preclude superluminal modes that are
known to emerge when $\gamma$ is negative. 

BIMM was discovered in a search for theories that generalize Maxwell
and BI in a manner that preserves the classical electromagnetic duality. 
The action is not invariant under the latter, as with Maxwell; it is better to invoke 
the Hamiltonian, which, modulo the Gauss constraint, may be written as
\begin{equation}
 \mathcal H_\gamma=\frac{1}{\bbe^2\ell^4}\left[\sqrt{1+\ell^4\cosh\gamma\;\mathbf w\cdot{\mathbf w}^*
 -\ell^4\sinh\gamma\;|\mathbf w\cdot\mathbf w|
 +\frac{\ell^8}{4}\,|{\mathbf w}\times\mathbf w^*|^2}\;-1\right] \ ,
 \label{bimm-hamiltonian}
\end{equation}
with 
\begin{equation}
 \mathbf w\equiv\bbe^2\mathbf D+i\mathbf B \ .
\end{equation}
The classical electromagnetic duality can be seen as a phase 
rotation $\mathbf w\rightarrow e^{i\varphi}\mathbf w$ that leaves $\cH_\gamma$ invariant. 
The same works for BI theory at $\gamma=0$, of course.

Starting from this BIMM theory coupled to a point charge $n$, one can compute
the self-force with the same tubular self-force formula (\ref{shift}).
Surprisingly, the results up to $\ell^3$ order are identical to BI results, upon
the following replacement of parameters, with $\bbe_\gamma$ and $\ell_\gamma$ given in (\ref{bimm-effective-parameters}),
\bea
q&\rightarrow & q_\gamma\equiv n\bbe_\gamma=e^{-\gamma/2}q\ ,\cr\cr
r_{\rm BI}&\rightarrow & r_\gamma\equiv\sqrt{\frac{\bbe_\gamma^2|n|}{4\pi}}\,\ell_\gamma
 =e^{-\gamma/4}r_{\rm BI} \ , \cr\cr
M_{\rm BI}&\rightarrow & M_\gamma\equiv\frac{q_\gamma^2\eta}{6\pi r_\gamma}
 =e^{-3\gamma/4}M_{\rm BI}\ .
 \label{bimm-redef}
\eea
Numerical computation suggests, however, that the agreement begins to fail 
at order $r_\gamma^4$.

This coincidental agreement to the cubic order in $\ell$ can be traced to how 
$J\sim \mathbf E\cdot \mathbf B$ plays little role in the lower orders. Recall
that, for both Maxwell and BI theories, $J$ vanishes for a charge in inertial 
motion and also for asymptotic waves. $J$ can be relevant in BI theory in principle 
via higher $\ell$ order corrections to the effective multipole
moments, but close inspection reveals that it does not enter
the final form of the self-force up to $r_{\rm BI}^3$ order. 

This motivates us to compare BI and BIMM at $J=0$, whereby we  find
\bea\label{no_J}
\mathcal L_\gamma\biggr\vert_{I<0;\, J=0} = \mathcal L_{\rm BI}\biggr\vert_{I<0; \,J=0}^{\bbe\rightarrow \bbe_\gamma;\,\ell\rightarrow \ell_\gamma}
\eea
with
\bea 
\bbe_\gamma^2\equiv\bbe^2e^{-\gamma}\ ,\qquad
 \ell_\gamma^4\equiv\ell^4e^{\gamma}\ .\label{bimm-effective-parameters}
\eea
Note that (\ref{bimm-effective-parameters}) combined with (\ref{no_J}) coupled to
a point charge $n$ suggests (\ref{bimm-redef}). In the weak field limit, for example,
the point  charge couples to the canonically normalized $\mathcal A_\mu/\bbe_\gamma$ with charge $q_\gamma$.
The correction due to $J\neq0 $ to (\ref{no_J})  occurs already at $\sim O(\ell^0)$
on the BIMM side but enters at $\sim O(\ell^4)$ on the BI side, so it is hardly obvious why 
our findings begin to differ at $\ell^4\sim \ell_\gamma^4$ scale. Nevertheless, one hint is
that the corrections to {\it on-shell} versions of (\ref{no_J}) at the core 
region actually start at $\ell^4 J^2$ on both sides.

\section{Radial Reductions for the Tubular Coefficients}
\label{app:tube-radial}

\subsection{Radial operators and core conditions}
For a radial profile ${\psi_l}(y)$ and a spherical harmonic satisfying
$\Delta_\Omega Y_{lm}=-l(l+1)Y_{lm}$, $l\ge1$, the divergence-free
radial and angular-gradient part of each $\mathbf{d}_N$ has the form
\begin{equation}
 \frac{l(l+1)\psi_l}{y^2}Y_{lm}\hat{\mathbf{r}}
                 +\frac{\psi_l'}y\nabla_\Omega Y_{lm} \ .
 \label{tube-general-radial}
\end{equation}
Here $\nabla_\Omega,\Delta_\Omega$ act on the unit sphere. The same
decomposition applies to the magnetic field. After applying $\mathsf M$
or $\mathsf K$, respectively, the curl has the angular factor
$(\hat{\mathbf{r}}\times\nabla_\Omega Y_{lm})/y$, with radial coefficients
\begin{equation}
 \begin{aligned}
 \mathscr D_l^E \psi_l&=\left(\frac{y^2\psi_l'}{\sqrt{1+y^4}}\right)'
 -\frac{l(l+1)y^4\psi_l}{(1+y^4)^{3/2}} \ ,\\
 \mathscr D_l^B \psi_l&=\left(\frac{\sqrt{1+y^4}}{y^2}\psi_l'\right)'
 -\frac{l(l+1)\psi_l}{\sqrt{1+y^4}} \ .
 \end{aligned}
 \label{tube-radial-operators}
\end{equation}
Projection of \eqref{tube-full-recursion} supplies the radial sources.
The radial components of the curl equations fix the fields along
$\hat{\mathbf{r}}\times\nabla_\Omega Y_{lm}$.

For radial lower-order leading displacements, the constitutive relation
and tangential Faraday equation give, under the differentiable core
assumptions of \eqref{tube-core-conditions},
\begin{equation}
 \lim_{y\to0}y^2[\mathbf{d}_N-(\mathbf{d}_N\cdot\hat{\mathbf{r}})\hat{\mathbf{r}}]
 =\lim_{y\to0}\mathbf{e}_N
 =-\lim_{y\to0}y\hat{\mathbf{r}}\times(\nabla_y\times\mathbf{e}_N)=0 \ .
 \label{tube-core-radial-consequence}
\end{equation}
Here $\mathbf{e}_N\cdot\hat{\mathbf{r}}\to0$ and $y\partial_y\mathbf{e}_N\to0$,
while neither Faraday source term has a $y^{-1}$ part.

\subsection{Linear profiles and retarded matching}
\label{app:tube-linear}
Integrating \eqref{tube-linear-main-recursion} introduces two constants
$\alpha_N,\beta_N$ for each profile $g_N$, giving
\begin{equation}
 g_N(y)=\frac{\alpha_N}y+\frac{\beta_N}y\int_0^y\sqrt{1+t^4}\,dt
 +\frac1y\int_0^y\sqrt{1+t^4}\int_0^t
 \frac{v g_{N-2}(v)}{\sqrt{1+v^4}}\,dv\,dt
 \quad(N\ge2) \ .
 \label{tube-linear-inverse}
\end{equation}

\paragraph{Core and exterior conditions.}
Both homogeneous modes of \eqref{tube-linear-inverse} also occur for $g_0,g_1$.
The mode $\alpha_N/y$ translates the static seed and changes its source,
\begin{equation}
 \nabla_y\cdot[-\alpha_N(\mathbf{a}^{(N)}\cdot\nabla_y)\mathbf{d}_0]
 =-4\pi\alpha_N(\mathbf{a}^{(N)}\cdot\nabla_y)\delta^{(3)}(\mathbf{y}) \ ,
 \label{tube-excluded-core-mode}
\end{equation}
so $\alpha_N=0$. The other mode gives an allowed $O(y^{-2})$
displacement. Its profile tends to $\beta_N=g_N(0)$, since the
particular integrals vanish at the core. In the overlap region,
$\delta g_N\sim\beta_N y^2/3$ gives
$\delta\mathbf{e}_{N+1}^{\rm lin}\to(2\beta_N/3)\mathbf{a}^{(N)}$.
The coefficient $\beta_N$ is fixed by matching this contribution,
together with the particular solution, to the exterior coefficient.

A large-$y$ electric term with radial power $m$ becomes, at fixed
physical $r$,
\begin{equation}
 \ell^{-2}r_{\rm BI}^N y^m
 =\frac{\bbe q}{4\pi}r_{\rm BI}^{N-2-m}r^m \ .
 \label{tube-inner-outer-powers}
\end{equation}

The retarded point-charge field, expressed in the comoving frame,
has the near-worldline expansion to first degree in the motion
\begin{equation}
 \left.\mathbf{E}_{\rm charge}^{\rm ret}\right|_{\text{degree }\le1}
 =\frac{\bbe q}{4\pi}\left[
 \frac{\hat{\mathbf{r}}}{r^2}
 -\frac{\mathbf{a}+(\mathbf{a}\cdot\hat{\mathbf{r}})\hat{\mathbf{r}}}{2r}
 +\frac23\dot{\mathbf{a}}+O(r)\right] \ .
 \label{tube-charge-matching}
\end{equation}
With \eqref{tube-field-series} and $r=r_{\rm BI}y$, a term
$r^{N-1}\mathbf{a}^{(N)}$ contributes at field order $N+1$.

\paragraph{Field order $N=1$.}
For the lapse source in \eqref{tube-linear-main-recursion},
the core-regular solution gives
\bea
 \mathbf{e}_1=\frac23(\beta_0+\eta)\mathbf{a}
 -\frac{\mathbf{a}+(\mathbf{a}\cdot\hat{\mathbf{r}})\hat{\mathbf{r}}}{2y}+O(y^{-3}) \ .
\eea
By \eqref{tube-inner-outer-powers}, the $y^0$ term would give an
exterior contribution proportional to $r_{\rm BI}^{-1}\mathbf{a}$.
The prescribed Maxwell limit at fixed $r$ excludes this term, so $\beta_0=-\eta$, giving
\begin{equation}
 g_0(y)=-\frac{\sqrt{1+y^4}}3\int_y^\infty\frac{dt}{\sqrt{1+t^4}}
 +\frac1{3y}\left[\left(\int_y^\infty\frac{dt}{\sqrt{1+t^4}}\right)^2-\eta^2\right]
 -\frac y6 \ ,
 \label{tube-g0}
\end{equation}
with core and exterior limits
\begin{equation}
 g_0(0)=-\eta \ ,\qquad
 g_0(y)=-\frac y2-\frac{\eta^2}{3y}+O(y^{-3}) \ .
 \label{tube-g0-ends}
\end{equation}
\Needspace{8\baselineskip}
Substituting the $-\eta^2/(3y)$ tail into
\eqref{tube-linear-profile}, with $\mathsf M_{ij}\to\delta_{ij}$ and
the physical normalization restored, gives
\bea
 -\frac{\bbe q}{4\pi}\frac{\eta^2r_{\rm BI}^2}{3r^3}
 \big[3(\mathbf{a}\cdot\hat{\mathbf{r}})\hat{\mathbf{r}}-\mathbf{a}\big] \ .
\eea
This fixes the order-$r_{\rm BI}^2$ exterior dipole amplitude.
At $N=1$ the magnetic sources vanish, and the boundary conditions
exclude a free magnetic field, so $\mathbf{b}_1=0$.

\paragraph{Field order $N=2$.}
The linear electric equation is $\mathscr D_1^Eg_1=0$.
With $\alpha_1=0$, the $\tfrac23\dot{\mathbf{a}}$ term in
\eqref{tube-charge-matching} fixes $\beta_1=1$, giving
\begin{equation}
 g_1(y)=\frac1y\int_0^y\sqrt{1+t^4}\,dt \ ,\qquad
 g_1(0)=1 \ ,\qquad
 g_1(y)=\frac{y^2}{3}+\frac{2\eta}{3y}+O(y^{-2}) \ .
 \label{tube-g1}
\end{equation}

\paragraph{Retarded exterior field.}
The $1/y$ tails in \eqref{tube-g0-ends} and \eqref{tube-g1}
have the dipole form generated by an auxiliary source displacement
$\delta\mathbf{z}$,
\begin{equation}
 -(\delta\mathbf{z}\cdot\nabla_{\mathbf x})
 \frac{\bbe q\hat{\mathbf{r}}}{4\pi r^2}
 =\frac{\bbe q}{4\pi r^3}
 [3(\delta\mathbf{z}\cdot\hat{\mathbf{r}})\hat{\mathbf{r}}-\delta\mathbf{z}] \ .
 \label{tube-tail-source-derivative}
\end{equation}
Variations along $a^\mu$ and $\dot a_i e_i^\mu$ reproduce the two tails with the
amplitudes in \eqref{tube-g0-ends} and \eqref{tube-g1}.
Applied to the conserved point-charge current, they generate the
electric dipole currents. At fixed physical radius and charge,
$\bbe^2\cP(\cF)-\cF=O(r_{\rm BI}^4)$, so the exterior equations at orders
two and three are Maxwell equations. Their retarded dipole fields
therefore follow by varying the point-charge retarded field. Within the local polynomial expansion,
dimensional analysis and parity allow only $r_{\rm BI}^2\mathbf a$ and
$r_{\rm BI}^3\dot{\mathbf a}$ as induced electric dipole amplitudes through
cubic length order. The charge field and these two dipole contributions
enter the exterior solution as
\begin{equation}
 \begin{aligned}
 \mathbf{E}_{\rm ext}^{\rm ret}[z]
 ={}&\mathbf{E}_{\rm charge}^{\rm ret}[z]
 -\frac{\eta^2r_{\rm BI}^2}{3}
 \left.\frac{d}{d\lambda}
 \mathbf{E}_{\rm charge}^{\rm ret}[z+\lambda a]\right|_{\lambda=0}\\
 &+\frac{2\eta r_{\rm BI}^3}{3}
 \left.\frac{d}{d\lambda}
 \mathbf{E}_{\rm charge}^{\rm ret}[z+\lambda\dot a_i e_i]\right|_{\lambda=0}
 +\cdots \ .
 \end{aligned}
 \label{tube-retarded-length-expansion}
\end{equation}
The auxiliary parameter $\lambda$ has dimension length squared in the first
variation and length cubed in the second. Both variations hold the observation
event and original comoving triad fixed and act on the source position,
tangent and retarded time determined by $(x-z(\tau_{\rm ret}))^2=0$ with
$x^0>z^0(\tau_{\rm ret})$. The small-$r$ expansion supplies both
linear and nonlinear matching data, with the full motion dependence
retained for the latter. Re-expansion at $r=r_{\rm BI}y$ in the
normalization of \eqref{tube-field-series} defines $\mathbf e_N^{\rm ret}$.
Below, $[\cdot]_{y^k}$ extracts the pure power $y^k$ from the large-$y$
expansion restricted to the stated angular and motion components.

\paragraph{Field orders $N=3,4,5$.}
At first order in the motion, the $r^0$ coefficient of the varied
physical electric field is $(\bbe q/4\pi)(2/3)d^3\delta\mathbf{z}/d\tau^3$.
With the normalization in \eqref{tube-field-series}, the two
variations give
\begin{equation}
 [\mathbf{e}_3^{\rm ret,lin}]_{y^0}=0 \ ,\qquad
 [\mathbf{e}_4^{\rm ret,lin}]_{y^0}=-\frac{2\eta^2}{9}\mathbf{a}^{(3)} \ ,\qquad
 [\mathbf{e}_5^{\rm ret,lin}]_{y^0}=\frac{4\eta}{9}\mathbf{a}^{(4)} \ .
 \label{tube-linear-dipole-matching}
\end{equation}
Matching \eqref{tube-linear-inverse} to these data fixes
$\beta_N=g_N(0)=\gamma_{N-1}$, $N=2,3,4$, as in \eqref{tube-gammas}.

\subsection{Nonlinear coefficients by field order}
\label{app:tube-nonlinear}
The core constants below use
the force normalization in \eqref{tube-core-trace} and the vector basis in
\eqref{tube-nonlinear-force}.

\paragraph{Field order $N=2$.}
\label{app:tube-second-fields}
Since $\mathbf{b}_1=0$, the source
$\bm{\mathcal E}_2+(\mathbf{a}\cdot\mathbf{y})\mathsf M\mathbf{d}_1$
is quadratic in $\mathbf{a}$, with $l=0,2$ from the trace and traceless
parts of $a_i a_j$. A divergence-free $l=0$ displacement is proportional
to $\hat{\mathbf r}/y^2$ and changes the charge, so its coefficient vanishes.
The remaining quadratic displacement defines $h(y)$,
\begin{equation}
 \mathbf{d}_2\big|_{a^2}
 =\frac{3h(y)}{y^2}[3(\mathbf{a}\cdot\hat{\mathbf{r}})^2-\mathbf{a}^2]\hat{\mathbf{r}}
 +\frac{3h'(y)}y(\mathbf{a}\cdot\hat{\mathbf{r}})
 [\mathbf{a}-(\mathbf{a}\cdot\hat{\mathbf{r}})\hat{\mathbf{r}}] \ .
 \label{tube-quadrupole-def}
\end{equation}
The $l=0$ radial source has zero curl. The $l=2$ projection gives
\begin{align}
 \mathscr D_2^Eh={}&-\frac{4y^4g_0^2}{(1+y^4)^{5/2}}
 +\frac{y^2(g_0')^2}{3{(1+y^4)^{3/2}}}+\frac{4y^5g_0}{3{(1+y^4)^{3/2}}}
\nonumber\\
 &-\left[-\frac{2y^2g_0g_0'}{3{(1+y^4)^{3/2}}}+\frac{y^3g_0'}{3{\sqrt{1+y^4}}}\right]' \ .
 \label{tube-quadrupole-ode}
\end{align}
The core-regular solution retains one homogeneous mode, which grows
as $y^3$. Since $\mathsf M$ approaches the identity and
$\bm{\mathcal E}_2=O(y^{-4})$, the full solution's radial electric field has
leading coefficient $3h/y^2$ multiplying the angular factor in
\eqref{tube-quadrupole-def}.

The retarded charge and dipole terms give
\begin{equation}
 \mathbf{e}_2^{\rm ret}\big|_{a^2}\cdot\hat{\mathbf{r}}
 =\left(\frac38+\frac{\eta^2}{6y^2}\right)
 [3(\mathbf{a}\cdot\hat{\mathbf{r}})^2-\mathbf{a}^2]+o(y^{-2}) \ .
 \label{tube-quadrupole-matching}
\end{equation}
Matching removes the $y^3$ term of the full $h$, which fixes the remaining
constant. The radial source already fixes the $y^2$ and $y^0$ terms,
giving $h=y^2/8+\eta^2/18+o(1)$, in agreement with
\eqref{tube-quadrupole-matching}.

\paragraph{Field order $N=3$.}
\label{app:tube-third-fields}
In $\mathbf{d}_3\cdot\hat{\mathbf{r}}$, terms proportional to $a_i a_j a_k$
have angular degrees $l=1,3$, while those proportional to
$a_i\dot a_j$ have $l=2$. The nonlinear radial magnetic component
is proportional to $(\mathbf{a}\times\dot{\mathbf{a}})\cdot\hat{\mathbf{r}}$
and has $l=1$. For the electric $\mathbf{a}^2\mathbf{a}$ contribution,
the core-regular solution and its retarded target give
\bea
 \begin{aligned}
 [\mathbf{e}_3]_{y^0}
 &=2\left(\frac{\delta_1}{3}+\frac{\eta^3}{27}
 -\frac{\pi}{36\eta}\right)\mathbf{a}^2\mathbf{a} \ ,\\
 [\mathbf{e}_3^{\rm ret}]_{y^0}&=0 \ .
 \end{aligned}
\eea
The constitutive remainder has no $y^0$ term. By \eqref{tube-inner-outer-powers}, the $y^0$ coefficient of the core-regular solution corresponds to
the absent order-$r_{\rm BI}$ exterior field. Matching gives
\eqref{tube-delta1}.
For the remaining radial components, the electric $l=3$, magnetic
$l=1$ and electric $l=2$ matching conditions are, respectively,
\bea
 \begin{aligned}
 [\mathbf{d}_3\cdot\hat{\mathbf{r}}]_{y^2}&=0 \ ,\qquad
 [\mathbf{b}_3\cdot\hat{\mathbf{r}}]_{y^0}=0 \ ,\\
 [\mathbf{d}_3\cdot\hat{\mathbf{r}}]_{y^1}
 &=-\frac23\left[3(\mathbf{a}\cdot\hat{\mathbf{r}})
 (\dot{\mathbf{a}}\cdot\hat{\mathbf{r}})-\mathbf{a}\cdot\dot{\mathbf{a}}\right] \ .
 \end{aligned}
\eea

\paragraph{Field order $N=4$.}
\label{app:tube-order4}
For the $l=1$ electric terms proportional to $\mathbf{a}^2\dot{\mathbf{a}}$
and $(\mathbf{a}\cdot\dot{\mathbf{a}})\mathbf{a}$, the core-regular solution gives
\begin{equation}
 [\mathbf{e}_4]_{y^0}
 =\frac23\left[
 \left(A_2-\frac{\eta^2}{3}\right)\mathbf{a}^2\dot{\mathbf{a}}
 +\left(B_2-\frac{\eta^2}{3}\right)(\mathbf{a}\cdot\dot{\mathbf{a}})\mathbf{a}\right] \ .
 \label{tube-order2-core-integrals}
\end{equation}
The $\eta^2/3$ terms come from the integrated source. The retarded target is
\begin{equation}
 [\mathbf{e}_4^{\rm ret}]_{y^0}
 =\frac{2\eta^2}{9}
 [\mathbf{a}^2\dot{\mathbf{a}}+(\mathbf{a}\cdot\dot{\mathbf{a}})\mathbf{a}] \ .
 \label{tube-order2-matched-profiles}
\end{equation}
Matching the two independent vector coefficients gives
$A_2=B_2=2\eta^2/3$.

\Needspace{11\baselineskip}
\paragraph{Field order $N=5$.}
\label{app:tube-order5}
For each of the five nonlinear vectors in \eqref{tube-nonlinear-force},
the electric displacement has the profile normalization of
\eqref{tube-linear-profile}, with that vector replacing $\mathbf a^{(N)}$.
Its core-regular profile contains $b_i g_1$ and a particular integral
of the lower-order source, vanishing at $y=0$.
Since $g_1(0)=1$ and $[g_1]_{y^2}=1/3$, the remaining constant contributes
$b_i$ to the force coefficient and $2b_i/3$ to the $y^0$ electric coefficient.
For these five nonlinear $l=1$ contributions, the retarded field
\eqref{tube-retarded-length-expansion} gives
\begin{equation}
 [\mathbf{e}_5^{\rm ret}]_{y^0}
 =-\frac{4\eta}{9}\left[
 \mathbf{a}^2\ddot{\mathbf{a}}+(\mathbf{a}\cdot\dot{\mathbf{a}})\dot{\mathbf{a}}\right] \ .
 \label{tube-retarded-coefficient-extraction}
\end{equation}
Matching the sum of the particular and homogeneous fields to these
coefficients determines $b_1,\ldots,b_5$, giving
\eqref{tube-b1}--\eqref{tube-b5}.

\end{document}